\documentclass[a4paper,fleqn]{cas-sc}

\usepackage[numbers]{natbib}

\usepackage{adjustbox}
\usepackage{bm}
\usepackage{xcolor}
\usepackage[many]{tcolorbox}

\newtheorem{thm}{Theorem}[section]
\newtheorem{teo}{Theorem}[section]
\newtheorem{cor}{Corollary}[thm]
\newtheorem{pro}[thm]{Proposition}

\newtheorem{lema}[thm]{Lemma}
\newproof{pf}{Proof}
\newproof{dem}{Proof}

\newtcolorbox{myboxi}[1][]{
  breakable,
  title=#1,
  colback=white,
  colbacktitle=white,
  coltitle=red,
  fonttitle=\bfseries,
  bottomrule=0pt,
  toprule=0pt,
  leftrule=3pt,
  rightrule=3pt,
  titlerule=0pt,
  arc=0pt,
  outer arc=0pt,
  colframe=red,
}

\makeatletter
\def\ps@pprintTitle{%
    \let\@oddhead\@empty
    \let\@evenhead\@empty
    \def\@oddfoot{\footnotesize\itshape
         {R Hern\'andez-Amador, JA Vallejo and Yu Vorobiev}}%
    \let\@evenfoot\@oddfoot
    }
\makeatother

\def\cm{\mathcal{C}^\infty (M)}
\def\xm{\mathcal{X}(M)}
\def\xmg{\mathcal{X}_G (M)}
\newcommand{\Ext}{\mathord{\adjustbox{valign=B,totalheight=.6\baselineskip}{$\bigwedge$}}}
\def\nnabla{\nabla \hskip-2.2mm \nabla}
\def\supernnabla{\nabla \hskip-2.0mm \nabla}

\newcommand{\bomega}{\boldsymbol{\omega}}
\newcommand{\blambda}{\boldsymbol{\lambda}}
\newcommand{\derm}[1]{\mathrm{Der}^{#1}\,\Omega(M)}
\newcommand{\produ}[3]{\left\langle #1,#2;#3\right\rangle}
\newcommand{\produc}[2]{\left\langle #1;#2\right\rangle}

\begin{document}
\let\WriteBookmarks\relax
\def\floatpagepagefraction{1}
\def\textpagefraction{.001}
\shorttitle{Symplectic scalar supercurvature on supermanifolds}
\shortauthors{R Hern\'andez-Amador, JA Vallejo and Yu Vorobiev}

\title[mode=title]{Symplectic scalar curvature on supermanifolds}                      

\author[1]{R Hern\'andez-Amador}[]
\ead{guadalupehernandez@correoa.uson.mx}
\address[1]{Departamento de Matem\'aticas, Universidad de Sonora,\\ 
Blvd L. Encinas y Rosales s/n Col. Centro, Ed. 3K-1\\
CP 83000 Hermosillo (Son) M\'exico.}

\author[2]{JA Vallejo}[orcid=0000-0002-9508-1549]
\cormark[1]
\ead{jvallejo@fc.uaslp.mx}
\ead[URL]{http://galia.fc.uaslp.mx/~jvallejo}
\address[2]{Facultad de Ciencias, Universidad Aut\'onoma de San Luis Potos\'i,\\
Av Chapultepec 1570 Col. Lomas del Pedregal\\
CP 78295 San Luis Potos\'i (SLP) M\'exico}

\author[1]{Yu Vorobiev}[]
\ead{yurimv@guaymas.uson.mx}

\cortext[cor1]{Corresponding author}


\begin{abstract}
We study the notion of symplectic scalar curvature on the supermanifold over an ordinary Fedosov manifold whose
structural sheaf is that of differential forms. In this purely geometric context, we introduce two families of
odd super-Fedosov structures, the first one is very general and uses a graded symmetric connection, leading to a
vanishing odd symplectic scalar curvature, while the second one is based on a graded non-symmetric connection and
has a non-trivial odd symplectic scalar curvature. As a simple example of the second case, we determine that curvature when the base Fedosov manifold is the torus.
\end{abstract}


\maketitle

\section{Introduction}

Let $M$ be a smooth manifold, $\xm$ the $\cm -$module of sections of its tangent bundle
(that is, the vector fields on $M$), and $\nabla:\xm\times\xm\to\xm$ a linear (Koszul)
connection. Using the canonical commutator of endomorphisms of a linear space and the
Lie bracket on $\xm$, the curvature of $\nabla$, $\mathrm{Curv}^\nabla$, is defined as
the tensor field $\mathrm{Curv}^\nabla\in\Omega^2 (M;\mathrm{End}\,TM)$ given by
\begin{equation}\label{eq0}
\mathrm{Curv}^\nabla (X,Y) = \left[\nabla_X ,\nabla_Y\right] -\nabla_{[X,Y]}\,.
\end{equation}

When $M$ supports an additional geometric structure, such as a Riemannian metric or a 
symplectic form, it is possible to pass from $\mathrm{Curv}^\nabla$ to a fully covariant
tensor of type $(0,4)$. Thus, if $B\in\mathcal{T}_2(M)$ is a non-degenerate field of
bilinear forms, we can define the Riemann $B-$curvature tensor of $\nabla$, $R^\nabla_B$,
by putting
\[
R^\nabla_B (X,Y,U,V)=B(\mathrm{Curv}^\nabla (X,Y)U,V)\,.
\]

When $B=g\in S_2 (M)$ is a pseudo-Riemannian metric ($2-$covariant, symmetric and 
non-degenerate), $R^\nabla_g$ is simply called the Riemann curvature tensor ; if
$B=\omega\in\Omega^2(M)$ is a symplectic form ($2-$form, non-degenerate and closed), 
$R^\nabla_\omega$ is called the symplectic curvature tensor. Particular cases of this setting 
are Riemannian manifolds with the compatible Levi-Civit\'a connection $\nabla g=0$,
and Fedosov manifolds, where $\nabla \omega =0$ instead \cite{Fed94,GRS98}.

A good deal of the information encoded in $R^\nabla_B$ can be recovered from another 
$2-$covariant tensor field obtained from it: the so-called $B-$Ricci tensor is defined as
the contraction given by the trace with respect to $B$:
\[
\mathrm{Ric}^\nabla_B (X,Y)=\mathrm{Tr}_B(Z\mapsto R^\nabla_B (Z,X,Z,Y))\,.
\]
A fundamental property of $\mathrm{Ric}^\nabla_B$ is its symmetry, no matter if $B$ is
symmetric or skew-symmetric:
\[
\mathrm{Ric}^\nabla_B (X,Y)=\mathrm{Ric}^\nabla_B (Y,X)\,.
\]
This property lies at the bottom of the different behaviour of $B=g$ a pseudo-Riemannian 
metric and  $B=\omega$ a symplectic form, when it comes to make a further contraction to 
define a scalar function related to the curvature: In the first case, we would get a
(generally) non-vanishing function called the Riemannian scalar curvature (or scalar
curvature, for short),
\[
\mathrm{Scal}^\nabla_g=\mathrm{Tr}_g (Z\mapsto \mathrm{Ric}^\nabla_g (Z,Z))\,,
\]
while in the second case we always get a vanishing symplectic scalar curvature,
\[
\mathrm{Scal}^\nabla_\omega=\mathrm{Tr}_\omega (Z\mapsto \mathrm{Ric}^\nabla_g (Z,Z))=0\,.
\]
Of course, this is due to the fact that the trace with respect to a skew-symmetric bilinear 
form of a symmetric one, is zero. Thus, there are no scalar invariants associated to the
symplectic curvature \cite{GRS98}.

In a series of interesting works, I. Batalin and K. Bering have pointed out that the 
situation is quite different in the category of supermanifolds \cite{BK08a,BK08b}. 
In it, there are even
symplectic forms and odd ones, the latter possessing the required symmetries to give a
non-vanishing odd symplectic scalar curvature. To explain this, let us recall that,
roughly speaking, in the Kostant-Leites-Manin approach a supermanifold can be thought as
just a usual manifold where the r\^ole of smooth functions is played by the sheaf of sections 
of the exterior bundle $\Ext E$ of some vector bundle $\pi:E\to M$ 
\cite{Kos77,Bat79,Lei80,Man97}. 
Thus,
supervector fields are derivations $\xmg =\mathrm{Der}_{\cm}\Ext E$, $1-$superforms are
the elements of the dual sheaf $\Omega^1_G (M)=\mathrm{Der}^\ast_{\cm}\Ext E$, and so on
\cite{SV09}.
Notice that all of these constructions inherite the $\mathbb{Z}-$grading of
$\Ext E=\oplus_{k\in\mathbb{Z}}\Ext^k E$, therefore, also the induced $\mathbb{Z}_2-$grading
which allows us to speak about the bosonic (or even-degree) and the fermionic (or odd-degree) 
sectors. A symplectic superform, thus, will be a mapping
\[
\bm{\omega} :\xmg\times\xmg\to\Gamma\Ext E\,,
\]
whose action on $D,D'\in\xmg$ will be denoted 
$\left\langle D,D';\bm{\omega}\right\rangle$ to stress that they form a 
$\Gamma\Ext E-$module, satisfying the graded analogs of the usual 
properties (see Section \ref{sec2} below). If $|D|,|D'|$ denote the respective degrees of the
supervector fields $D,D'$, we will have an even symplectic form when
\begin{equation}\label{eq1}
\left\langle D',D;\bm{\omega}\right\rangle =
-(-1)^{|D||D'|}\left\langle D,D';\bm{\omega}\right\rangle\,,
\end{equation}
and an odd one when
\begin{equation}\label{eq2}
\left\langle D',D;\bm{\omega}\right\rangle =
-(-1)^{(|D|-1)(|D'|-1)}\left\langle D,D';\bm{\omega}\right\rangle\,,
\end{equation}

In this setting, a graded connection is a mapping 
$\nnabla :\mathrm{Der}\Ext E\times\mathrm{Der}\Ext E\to\mathrm{Der}\Ext E$ also satisfying
the graded version of the usual properties. As in the classical (non-graded) case, the action
of $\nnabla$ can be extended to any tensor superfield in the form of a covariant 
superderivative. If $(M,\Gamma\Ext E)$ is a supermanifold, the triple 
$((M,\Gamma\Ext E),\bm{\omega},\nnabla)$ is a Fedosov supermanifold when 
$\nnabla\bm{\omega}=0$. In dealing with symplectic supercurvatures, as in the non-graded 
case, particular attention will be paid to compatible connections, that is, to Fedosov
supermanifolds \cite{AL09,MMV09,Val12}.
The Riemann supercurvature of a graded connection $\nnabla$ can be defined as usual, with
the aid of the graded canonical commutator of endomorphisms of superalgebras,
$[\![\cdot,\cdot ]\!]$:
\[
\mathrm{Curv}^{\nnabla} (D,D')=[\![\nnabla_D,\nnabla_{D'}]\!] -\nnabla_{[\![D,D']\!]}\,.
\]

Similarly, we can define the graded Ricci tensor associated to a given graded connection
$\nnabla$, $\mathrm{Ric}^{\nnabla}$, which turns out to always be even-symmetric, in the 
sense that
\begin{equation}\label{eq3}
\mathrm{Ric}^{\nnabla}(D,D')=(-1)^{|D||D'|}\mathrm{sRic}^{\nnabla}(D',D)\,.
\end{equation}
The contraction of $\mathrm{Ric}^{\nnabla}$ with an even symplectic form \eqref{eq1}
will give zero as in the non-graded category, but the surprising case is the contraction
with an odd symplectic form \eqref{eq2}, as now the reasoning based on just symmetry 
properties fails to imply that the result must be zero.

Assuming that this contraction is non-trivial, Batalin and Bering went on to relate the
resulting odd symplectic scalar curvature with the eigenvalues of the odd Laplacian 
\cite{BK08a,BK08b}.
However, they did not touch the issue of non-triviality, so at the present 
we do not have a complete answer to the question `when is the odd symplectic scalar curvature 
non zero?' Our aim in this work is to present some results in this direction, along with
explicit examples, in a particular class geometric supermanifolds, those whose structural 
sheaf $\Gamma\Ext E$ is precisely the sheaf of sections of the exterior bundle of the base
manifold. These supermanifolds, of the form $(M,\Omega (M))$ will be named Koszul-Cartan
supermanifolds here \cite{Kos19}. From a purely physical point of view, they are somewhat restricted,
because if the dimension of the base manifold is $m=\dim M$, the fermionic content of
any physical theory constructed upon them contains a limited number of fields. But this
lack of generality does not mean that this class of supermanifolds is uninteresting, as we
argue next: It can be proved that up to second-order depth, the structure of graded 
symplectic structures on the supermanifold $(M,\Omega (M))$ is the following \cite{Mon92},
\begin{enumerate}
\item\mbox{}
\begin{equation}\label{eq4}
\bm{\omega}_{\tilde{\omega},g}=\begin{pmatrix}
\tilde{w} &  \\
			   & g
\end{pmatrix}\quad\mbox{ if }\bm{\omega}\mbox{ is even,}
\end{equation}
where $\tilde{w}\in\Omega^2(M)$ is an ordinary symplectic form and $g$ is a
pseudo-Riemannian metric on the base manifold $M$.
\item\mbox{}
\begin{equation}\label{eq5}
\bm{\omega}_H=\begin{pmatrix}
  &  -H \\
H & 
\end{pmatrix}\quad\mbox{ if }\bm{\omega}\mbox{ is odd,}
\end{equation}
where $H:TM\to T^*M$ is a bundle isomorphism.
\end{enumerate}
These local expressions must be understood in the sense that given any local basis of vector
fields $X_i\in\xm$ (with $1\leq i\leq n=\dim M$), then, for any linear connection $\nabla$ on
$TM$, $\left\lbrace\nabla_{X_j},i_{X_j}\right\rbrace^n_{j=1}$ (where $i_X$ denotes the
insertion operator) is a basis for the sheaf (of finitely generated, locally free
$\cm -$modules) $\mathrm{Der}_{\cm} \Omega(M)$. Thus, any
symplectic superform can be characterized by giving the supermatrices
\[
\begin{pmatrix}
\left\langle\nabla_{X_j},\nabla_{X_k};\bm{\omega}\right\rangle &  &
\left\langle\nabla_{X_j},i_{X_k};\bm{\omega}\right\rangle \\[10pt] 
\left\langle i_{X_j},\nabla_{X_k};\bm{\omega}\right\rangle &  &
\left\langle i_{X_j},i_{X_k};\bm{\omega}\right\rangle
\end{pmatrix}\,,
\]
where the matrix elements do not depend on the chosen connection $\nabla$ (this is in marked
contrast to the case of graded connections, whose local expressions depend crucially on 
$\nabla$).
Of course, a particular class of odd symplectic forms \eqref{eq5} is obtained by considering
$w\in \Omega^2(M)$ a symplectic structure on $M$ and the induced musical isomporphism
$H=\flat$ (we will continue to denote it by $w$ in a slight abuse of notation), resulting
in the local expression
\begin{equation}\label{eq6}
\bm{\omega}_{w}=\begin{pmatrix}
  &  -w \\
w & 
\end{pmatrix}\,.
\end{equation}

We can always choose the connection $\nabla$ to be symplectic, that is, such that the $2-$form $w$ is parallel, $\nabla w=0$.
Thus, by considering the Koszul-Cartan supermanifold we have a canonical way of relating non-graded Fedosov structures
(on the base manifold) and super-Fedosov ones, through \eqref{eq6}.

In this work we carry on a mathematical study
of the properties of the symplectic curvature on Koszul-Cartan supermanifolds, obtaining
the following main results:
\begin{enumerate}
\item It is possible to construct a class of odd-symplectic supermanifolds 
$((M,\Omega(M)),\bm{\omega}_H)$ starting from a usual manifold $M$ endowed with an
isomorphism $H:TM\to T^*M$. If, moreover, a symmetric linear connection $\nabla$ on $M$
is also considered, from $(M,\nabla,H)$ it is possible to define a quite general class of Fedosov
supermanifolds $((M,\Omega(M)),\nnabla,\bm{\omega}_H)$, with $\nnabla$ graded symmetric, whose elements are completely
determined by a set of six tensor fields on the base manifold $M$ (Section \ref{sec-fedosov}).
\item If $\nabla H=0$, then the odd symplectic scalar curvature of any member of the above 
family is zero. In a certain sense, this kind of a no-go theorem explains why it is so
difficult to find explicit examples of non-trivial symplectic scalar supercurvatures, as the
natural thing to do is to consider parallel structures in trying to construct them (Section \ref{sec4}).
\item If the base is a Fedosov manifold $(M,\nabla,w)$, and $H=\flat:TM\to T^*M$ is the isomorphism induced by 
the symplectic form $w$, then it is possible to construct a family of Fedosov supermanifolds
$((M,\Omega(M)),\nnabla,\bm{\omega}_g)$, with a \emph{graded non-symmetric connection} $\nnabla$ such that,
in general, has non-trivial odd symplectic scalar curvature (Section \ref{sec5}). In order to show the geometric
origin of this curvature, we analyze what is perhaps the simplest example possible, built on the two-dimensional
torus (Section \ref{lastsection}). 
\end{enumerate}

Previous to that, in Section \ref{sec2} we briefly recall the main results about the geometry of supermanifolds,
establishing the notation and conventions used in the paper.
Appendix \ref{appendix} collects some basic facts about Fedosov structures on the torus that will be needed
in the text.

\section{Geometry of supermanifolds}\label{sec2}

All along the paper, $M$ will be a smooth manifold with $\dim M=n$. A real supermanifold is a ringed space $(M,\mathcal{A})$, 
where $\mathcal{A}$ is a sheaf of $\mathbb{Z}_2 -$graded commutative $\mathbb{R}-$algebras (the structural sheaf) such that:
\begin{enumerate}[(a)]
\item If $\mathcal{N}$ denotes the sheaf of nilpotents of $\mathcal{A}$, then $\mathcal{A}/\mathcal{N}$ induces on $M$ the structure of a differential manifold (the base).
\item The subsheaf $\mathcal{N}/\mathcal{N}^2$ is a locally free sheaf of modules, with $\mathcal{A}$ \emph{locally} isomorphic to the exterior sheaf $\bigwedge \left( \mathcal{N}/\mathcal{N}^2 \right)$.
\end{enumerate}

A theorem of M. Batchelor \cite{Bat79} states that for a real graded manifold, letting
$\mathcal{E}=\mathcal{N}/\mathcal{N}^2$, we have a sheaves isomorphism
\begin{equation}\label{splitting}
\mathcal{A}\simeq \bigwedge \mathcal{E}
\end{equation}
\noindent not only locally, but globally. However, this sheaf isomorphism is \emph{not} canonical. When the identification 
\eqref{splitting} is made, it is said that the supermanifold $(M,\mathcal{A})$ is given in \emph{split} form. 
An interesting result, 
due to Koszul \cite{Kos-94}, is that a graded manifold splits if and only if $\mathbb{Z}_2 -$graded connections exist (see 
below for the definition and main properties of graded connections). 
In this paper, we will consider the Cartan-Koszul supermanifold $(M,\Omega (M))$ \cite{Kos19}, for which the structural sheaf 
is that of differential forms on the manifold $M$, that is, 
$\mathcal{A}(U)=\Omega (U)=\bigoplus_{p\in \mathbb{Z}}\Omega^p (U)$.

As for ordinary $\mathcal{C}^\infty-$manifolds, (super) vector fields on the supermanifold 
$(M,\Omega (M))$ are now the graded derivations $\mathrm{Der}\,\Omega (M)$ 
(such as the exterior differential $\mathrm{d}$, which has degree 
$|\mathrm{d}|=1$, 
the Lie derivative $\mathcal{L}_X$, which has degree $|\mathcal{L}_X|=0$, 
or the insertion of a vector field 
$i_X$, which has degree $|i_X|=-1$). Given a linear connection $\nabla$ on $M$, 
derivations of the form  $\nabla_X$, $i_X$ generate the $\Omega (M)-$module $\derm{}$, according to a well-known theorem
by Fr\"olicher and Nijenhuis. The notation $\derm{}=\mathcal{X}^G (M)$ will also be used.

\subsection{Graded symplectic forms}

The (super) differential $1-$forms on 
$(M,\Omega (M))$ are defined as the duals $\mathrm{Der}^\ast \Omega (M)$, and
$k-$forms are defined by taking exterior products as usual, and noting that they
are \emph{bigraded} objects; if, for instance, 
$\boldsymbol{\omega}\in\Omega^2(M,\Omega (M))$ 
(that is the way of denoting the space
of $2-$superforms, sometimes we will use a notation such as $\Omega^2_G(M)$), its action on two supervector fields $D,D'\in\mathrm{Der}\,\Omega(M)$ will be denoted $\left\langle D,D';\boldsymbol{\omega}\right\rangle$, 
a notation well adapted to the fact that $\mathrm{Der}\,\Omega(M)$ is considered 
here as a left $\Omega(M)-$module and $\Omega^2(M,\Omega (M))$ as a right one. 
Other objects such as the graded exterior differential can be defined as in the 
classical setting, but taking into account the $\mathbb{Z}_2-$degree (see \cite{Val12} for details). 
Thus, if $\alpha\in\Omega^0 (M,\Omega(M))$, its graded differential $\boldsymbol{\mathrm{d}}$ 
is given by
$\left\langle D;\boldsymbol{\mathrm{d}}\alpha\right\rangle =D(\alpha)$, and if
$\boldsymbol{\beta} \in \Omega^1 (M,\Omega(M))$, we have a $2-$form 
$\boldsymbol{\mathrm{d}}\boldsymbol{\beta} \in \Omega^2 (M,\Omega(M))$ whose action is 
given by
\[
\left\langle D,D';\boldsymbol{\mathrm{d}}\boldsymbol{\beta}\right\rangle
=D(\left\langle D';\boldsymbol{\beta}\right\rangle)
-(-1)^{|D||D'|}D' (\left\langle D;\boldsymbol{\beta}\right\rangle)
-\left\langle [D,D'];\boldsymbol{\beta}\right\rangle\,,
\]
where $|D|$ denotes the degree of the derivation $D$.

A graded symplectic form is a closed graded $2-$form $\boldsymbol{\omega} \in \Omega^2_G (M)$ that is non singular, that is, the $\Omega(M)-$linear map it induces
\begin{center}
\begin{tabular}{rcl}
$\mathcal{X}^G (M)$ & $\rightarrow$ & $\Omega^1_G (M)$ \\
$D$ & $\mapsto$ & $\iota_D \boldsymbol{\omega}$, 
\end{tabular}
\end{center}
is an isomorphism.
Notice that there are two classes of supersymplectic forms: the even ones (for which $|\bomega|$ is even) act in such a way that, in terms of the induced $\mathbb{Z}_2-$degree,
\[
|\produ{D}{D'}{\bomega}|=|D|+|D'|
\]
while the odd symplectic forms (for which $|\bomega|$ is odd) satisfy
\[
|\produ{D}{D'}{\bomega}|=|D|+|D'|+1\,.
\]

We will restrict our attention to a particular class of odd symplectic forms, constructed as follows.
Consider a vector bundle isomorphim $H:TM\to T^*M$, and
define an odd $1-$form, $\blambda_H$, by its action on basic derivations:
\begin{align*}
&\produc{\nabla_X}{\blambda_H}= H(X)\\
&\produc{i_X}{\blambda_H}= 0\,.
\end{align*}
(notice that this action is actually independent of $\nabla$). Then, define 
$\bomega_H$ by $\bomega_H =\boldsymbol{\mathrm{d}}\blambda_H$. Thus,
the action of $\bomega_H$ on basic derivations now reads
\begin{align}\label{bomegah}
&\produ{\nabla_X}{\nabla_Y}{\bomega_H}= (\nabla_X H)Y-(\nabla_Y H)X\nonumber \\
& \produ{\nabla_X}{i_Y}{\bomega_H}= -H(X)(Y)\nonumber \\
& \produ{i_X}{\nabla_Y}{\bomega_H}= H(Y)(X) \\
& \produ{i_X}{i_Y}{\bomega_H}= 0\nonumber\,.
\end{align}

\subsection{Graded connections}
Along with the notion of graded symplectic forms, in order to define Fedosov supermanifolds we will need a generalization of
the idea of compatible connections.

A graded connection on a graded manifold $(M,\mathcal{A})$ is a mapping
\begin{center}
\begin{tabular}{rcl}
$\nnabla :\mathcal{X}^G (M)\times \mathcal{X}^G (M)$ & $\rightarrow$ & $\mathcal{X}^G (M)$ \\
$(D_1 ,D_2 )$ & $\mapsto$ & $\nnabla_{D_1} D_2$ 
\end{tabular}
\end{center}
such that, for all $D_1 ,D_2 ,D_3 \in\mathcal{X}^G (M)$ and $\alpha \in \mathcal{A}$,
\begin{enumerate}[(1)]
\item $\nnabla_{D_1} (D_2 +D_3 )=\nnabla_{D_1} D_2 +\nnabla_{D_1} D_3$,
\item $\nnabla_{(D_1 +D_2 )}D_3 =\nnabla_{D_1} D_3 +\nnabla_{D_2} D_3$,
\item $\nnabla_{\alpha D_1} D_2 =(-1)^{|\alpha ||\nnabla |}\alpha \nnabla_{D_1} D_2$,
\item $\nnabla_{D_1}(\alpha D_2) =D_1 (\alpha )D_2 +(-1)^{|\alpha |(|D_1 |+|\nnabla |)}\alpha \nnabla_{D_1} D_2\,.$
\end{enumerate}

A graded connection in $(M,\mathcal{A})$ is called $\mathbb{Z}-$homogeneous of degree $|\nnabla |$ if for any pair of 
homogeneous derivations $D_1 ,D_2 \in \mathcal{X}^G (M)$, $\nnabla_{D_1} D_2 \in\mathcal{X}^G (M)$ is homogeneous and 
$| \nnabla_{D_1} D_2 |=|D_1 |+|D_2 |+|\nnabla |$. If
$|\nnabla |\equiv 0\,\mathrm{mod}\,2$, it is said that $\nnabla$ is even, and odd if $|\nnabla |\equiv 1\,\mathrm{mod}\,2$ 
(in either case, the connection is said to be $\mathbb{Z}_2 -$graded). From now on, unless otherwise explicitly stated, 
we will consider only even connections.

The torsion of $\nnabla$ is the mapping $\mathrm{Tor}^{\nnabla}:\mathcal{X}^G (M)\times \mathcal{X}^G (M)\to \mathcal{X}^G (M)$ 
given by
\[
\mathrm{Tor}^{\nnabla}(D_1 ,D_2 )=\nnabla_{D_1} D_2 -(-1)^{|D_1 ||D_2 |}\nnabla_{D_2} D_1 -[D_1 ,D_2 ]\,,
\] 
and the graded curvature $\mathrm{Curv}^{\nnabla}$ is given by
\[
\mathrm{Curv}^{\nnabla}(D_1 ,D_2 )D_3 =[\nnabla_{D_1}, \nnabla_{D_1}]D_3 -\nnabla_{[D_1 ,D_2]}D_3 \,,
\]
for all $D_1 ,D_2 ,D_3 \in \mathcal{X}^G (M)$ (so it is a graded tensor of type $(3,1)$). As in the non-graded case, we
say that $\nnabla$ is \emph{symmetric} if its torsion vanishes.

Notice that $\nnabla_{D} $ can be extended as an operator to all of $\Omega^G (M)$; for example, if $\boldsymbol{\tau}$ is a 
graded $2-$form, then $\nnabla_{D} \boldsymbol{\tau}$ is again a $2-$form, given by its action on homogeneous derivations 
$D_1 ,D_2$ (of respective $\mathbb{Z}-$degrees $|D_1 |,|D_2 |$):
\begin{equation*}
 \langle D_1,D_2;\nnabla_D \tau\rangle := 
 (-1)^{|D|(|D_1|+|D_2|)} \left\{D(\langle D_1,D_2;\tau \rangle) - 
 \langle \nnabla_D D_1,D_2;\tau\rangle \right.  \left.- 
 (-1)^{|D||D_1|}\langle D_1,\nnabla_D D_2;\tau\rangle \right\},
\end{equation*}
where we have used the notation $\langle D_1,D_2;\nnabla_D\boldsymbol{\tau}\rangle$ to stress the right
$\Omega (M)-$module structure.

The graded curvature has the same algebraic properties than its non-graded counterpart. The most important, undoubtedly, 
is the Bianchi identity (whose proof is exactly the same as in the non-graded case, just keeping track of the signs).

\begin{pro}
Let $\nnabla$ be a graded connection on $(M,\Omega(M))$. Then, the graded cyclic sum of its graded curvature with respect its 
arguments vanishes:
\[
\mathrm{Curv}^{\nnabla}(D_1 ,D_2 )D_3 + (-1)^{|D_1|(|D_2|+|D_3|)} \mathrm{Curv}^{\nnabla}(D_2 ,D_3 )D_1 + (-1)^{|D_3|(|D_1|+|D_2|)} \mathrm{Curv}^{\nnabla}(D_3 ,D_1 )D_2=0\,.
\]
\end{pro}

Once a linear connection $\nabla$ on $M$ is chosen, a superconnection is characterized 
by a set of tensor fields giving its action on basic derivations,
\begin{equation}\label{nnabla}
\begin{split}
&\nnabla_{\nabla_X}\nabla_Y=\nabla_{\nabla_XY+K_0(X,Y)} + i_{L_0(X,Y)}\\[3pt]
&\nnabla_{\nabla_X}i_Y =\nabla_{K_1(X,Y)} + i_{\nabla_X Y+L_1(X,Y)}\\[3pt]
&\nnabla_{i_X}\nabla_Y=\nabla_{K_2(X,Y)} + i_{L_2(X,Y)}\\[3pt]
&\nnabla_{i_X}i_Y =\nabla_{K_3(X,Y)} + i_{L_3(X,Y)}\,,
\end{split}
\end{equation}
where $K_i:\Gamma TM\otimes \Gamma TM\to \Gamma \Lambda^{\phi(i)}T^*M\otimes\Gamma TM$, and 
$L_i :\Gamma TM\otimes \Gamma TM\to \Gamma \Lambda^{\psi(i+1)}T^*M\otimes\Gamma TM$, 
for $i\in\{0,1,2,3\}$, $\phi$ and $\psi$ being functions that depend on the $\mathbb{Z}-$degree of $\nnabla$.
The following result put some conditions on these tensors when the connection is symmetric.

\begin{thm}\cite{MMV09}\label{thm21}
Let $\nabla$ be a symmetric linear connection on $M$. A superconnection 
$\nnabla$ on $(M,\Omega (M))$ is symmetric if and only if, for any $X,Y\in\mathcal{X}(M)$ the following are satisfied:
\begin{equation}\label{torsionless-connection}
\begin{array}{ll}
K_0(X,Y) = K_0(Y,X) & L_0(X,Y) =
L_0(Y,X)+ \mathrm{Curv}^\nabla(X,Y)\\
K_1(X,Y) = K_2(Y,X) & L_1(X,Y) = L_2(Y,X)\\
K_3(X,Y) = - K_3(Y,X) & L_3(X,Y) = - L_3(Y,X)\,.
\end{array}
\end{equation}
\end{thm}

\subsection{Graded curvature and Ricci tensors. Scalar supercurvature}

For $\nnabla$ a graded connection on $(M,\Omega(M))$ and $\bm\tau$ a graded covariant $2-$tensor, the graded curvature tensor
of $\nnabla$ with respect to $\bm\tau$ is defined as the mapping
$R^{\nnabla}_{\bm\tau}:\mathcal X^G(M) \times \mathcal X^G(M) \times\mathcal X^G(M) \times\mathcal X^G(M) \to \Omega(M)$
given by
\[
\langle D_1,D_2,D_3,D_4;R^{\nnabla}_{\bm\tau}\rangle
 =\langle \, \mathrm{Curv}^{\nnabla}(D_1,D_2)D_3),D_4 \,; \bm\tau\rangle\,.
\]

Notice that if $\bm\tau$ happens to be homogeneous, as a consequence of $\nnabla$ being even, the graded curvature
tensor is homogeneous of degree 
$|R^{\nnabla}_{\bm\tau}|=|\bm\tau|$.

The following result, whose proof is just a straightforward computation, guarantees that $R^{\nnabla}_{\bm\tau}$
has the required multilinearity property of a tensor, along with an important skew-symmetry property.

\begin{pro} \label{propsTensCurvTau}
For any graded convariant $2-$tensor $\bm\tau$, the tensor $R^{\nnabla}_{\bm\tau}$ satisfies:
 \begin{enumerate}
 \item If $\nnabla$ is even, $R^{\nnabla}_{\bm\tau}$ is a graded covariant $4-$tensor (that is,
 graded $\Omega (M)-$multilinear in any of its arguments).
 \item Is graded skew-symmetric in its two first arguments, that is, 
 \[
 \langle D_1,D_2,D_3,D_4;R^{\nnabla}_{\bm\tau}\rangle
 =-(-1)^{|D_1||D_2|} \langle D_2,D_1,D_3,D_4;R^{\nnabla}_{\bm\tau}\rangle
 \]
\end{enumerate}
\end{pro}

If $\nnabla$ is given by the tensor fields
$\{K_i,L_i\}$ as in \eqref{nnabla}, a lengthy but straightforward computation, using the Fr\"olicher-Nijenhuis decompositions
\[
[\nabla_X,\nabla_Y]=\nabla_{[X,Y]}+i_{\mathrm{Curv}^\nabla(X,Y)}
\]
and
\[
[\nabla_X,i_Y]=i_{\nabla_XY}\,,
\]
proves the following result about the structure of its curvature. 
\begin{pro}\label{pro24}
With the preceding notations, the graded curvature of a graded connection $\nnabla$ is determined by
\begin{equation}\label{curvatura_nnabla_ocho}
 \begin{split}
 \vspace{2mm}
 &\mathrm{Curv}^{\nnabla}(\nabla_X,\nabla_Y)\nabla_Z = \nabla_{A_1(X,Y,Z)} + i_{B_1(X,Y,Z)}  \\
 \vspace{2mm}
 &\mathrm{Curv}^{\nnabla}(\nabla_X,\nabla_Y)i_Z = \nabla_{A_2(X,Y,Z)} + i_{B_2(X,Y,Z)}  \\
 \vspace{2mm}
 &\mathrm{Curv}^{\nnabla}(\nabla_X,i_Y)\nabla_Z  = \nabla_{A_3(X,Y,Z)} + i_{B_3(X,Y,Z)}  
                                             = -\mathrm{Curv}^{\nnabla}(i_Y, \nabla_X)\nabla_Z \\
 \vspace{2mm}
 &\mathrm{Curv}^{\nnabla}(\nabla_X,i_Y)i_Z  = \nabla_{A_4(X,Y,Z)} + i_{B_4(X,Y,Z)}  
                                                  = -\mathrm{Curv}^{\nnabla}( i_Y, \nabla_X)i_Z \\ 
 \vspace{2mm}
 &\mathrm{Curv}^{\nnabla}(i_X,i_Y)\nabla_Z  = \nabla_{A_5(X,Y,Z)} +i_{B_5(X,Y,Z)}\\ 
 &\mathrm{Curv}^{\nnabla}(i_X,i_Y)i_Z = \nabla_{A_6(X,Y,Z)} +i_{B_6(X,Y,Z)}\,,
 \end{split}
 \end{equation}
where $X,Y,Z\in\mathcal{X}(M)$ are arbitrary, and
\begin{eqnarray}
A_1(X,Y,Z)&=& \,\,\,\nabla_X K_0(Y,Z) + K_0(X,\nabla_YZ + K_0(Y,Z))  + K_1(X,L_0(Y,Z)) \nonumber\\
           && -\nabla_Y K_0(X,Z)  - K_0(Y,\nabla_XZ + K_0(X,Z)) - K_1(Y,L_0(X,Z))\nonumber \\
           && + R^\nabla(X,Y)Z - K_0([X,Y],Z) - K_2(R^\nabla(X,Y),Z)  \nonumber\\        
B_1(X,Y,Z)&=&  \,\,\,\nabla_X L_0(Y,Z) + L_0(X,\nabla_YZ + K_0(Y,Z)) + L_1(X,L_0(Y,Z))  \nonumber\\
           &&  -\nabla_Y L_0(X,Z) - L_0(Y,\nabla_XZ + K_0(X,Z))  -L_1(Y,L_0(X,Z)\nonumber) \\
           &&  - L_0([X,Y],Z)- L_2(R^\nabla(X,Y),Z)\nonumber\\
A_2(X,Y,Z)&=& K_0(X,K_1(Y,Z)) + \nabla_X(K_1(Y,Z)) + K_1(X,\nabla_YZ + L_1(Y,Z))\nonumber\\ 
           && -K_0(Y,K_1(X,Z))  -\nabla_Y(K_1(X,Z)) - K_1(Y,\nabla_XZ + L_1(X,Z))\nonumber\\
           && -K_1([X,Y],Z)-K_3(R^\nabla(X,Y),Z) \nonumber\\
B_2(X,Y,Z) &=& L_0(X,K_1(Y,Z)) + \nabla_X L_1(Y,Z) + L_1(X,\nabla_YZ + L_1(Y,Z))\nonumber\\
            && -L_0(Y,K_1(X,Z)) -\nabla_Y L_1(X,Z) - L_1(Y,\nabla_XZ + L_1(X,Z)) \nonumber \\ 
            && + R^\nabla(X,Y)Z - L_1([X,Y],Z) - L_3(R^\nabla(X,Y),Z)\nonumber\\
A_3(X,Y,Z)&=& K_0(X,K_2(Y,Z)) + K_1(X,L_2(Y,Z)) + \nabla_X(K_2(Y,Z)) - K_2(Y,\nabla_XZ + K_0(X,Z)) \nonumber\\
            && - K_2(\nabla_XY,Z)-K_3(Y,L_0(X,Z))\nonumber\\
B_3(X,Y,Z) &=& L_0(X,K_2(Y,Z)) + L_1(X,L_2(Y,Z)) + \nabla_X L_2(Y,Z) - L_2(Y, \nabla_XZ + K_0(X,Z)) \nonumber\\
            && - L_2(\nabla_XY,Z) - L_3(Y,L_0(X,Z))\nonumber\\
A_4(X,Y,Z)&=& K_0(X,K_3(Y,Z))+K_1(X,L_3(Y,Z)) -K_2(Y,K_1(X,Z)) + \nabla_X K_3(Y,Z) \nonumber\\
           && - K_3(Y,\nabla_XZ+L_1(X,Z))-K_3(\nabla_XY,Z)\nonumber\\
B_4(X,Y,Z) &=& L_0(X,K_3(Y,Z)) +L_1(X,L_3(Y,Z)) -L_2(Y,K_1(X,Z)) + \nabla_X L_3(Y,Z) \nonumber\\
           &&- L_3(Y,\nabla_XZ+L_1(X,Z))-L_3(\nabla_XY,Z)\nonumber\\
A_5(X,Y,Z) &=& K_2(X,K_2(Y,Z))  + K_3(X,L_2(Y,Z)) \nonumber\\
             && + K_2(Y,K_2(X,Z)) + K_3(Y,L_2(X,Z))\nonumber\\
B_5(X,Y,Z) &=& L_2(X,K_2(Y,Z)) + L_3(X,L_2(Y,Z))\nonumber\\
            &&+ L_2(Y,K_2(X,Z))  + L_3(Y,L_2(X,Z))\nonumber\\
A_6(X,Y,Z) &=& K_2(X,K_3(Y,Z)) + K_3(X,L_3(Y,Z))\nonumber\\
            && +K_2(Y,K_3(X,Z)) + K_3(Y,L_3(X,Z))\nonumber\\
B_6(X,Y,Z) &=& L_2(X,K_3(Y,Z)) + L_3(X,L_3(Y,Z))\nonumber\\
            && L_2(Y,K_3(X,Z)) + L_3(Y,L_3(X,Z)). \nonumber
\end{eqnarray}
\end{pro}

Now, as the curvature $\mathrm{Curv}^{\nnabla}$ is even, for each pair $D_1,D_2\in\mathcal X^G(M)$ we can define the following
homogeneous operators of degree $|D_1|+|D_2|$:
\begin{enumerate}
 \item $\langle \cdot,D_1,D_2; \mathrm{Curv}^{\nnabla}\rangle :D\mapsto \langle D,D_1,D_2;\mathrm{Curv}^{\nnabla}\rangle$
 \item $\langle D_1,\cdot,D_2;\mathrm{Curv}^{\nnabla}\rangle :D\mapsto \langle D_1,D,D_2;\mathrm{Curv}^{\nnabla}\rangle$ 
 \item $\langle D_1,D_2,\cdot;\mathrm{Curv}^{\nnabla}\rangle :D\mapsto \langle D_1,D_2,D;\mathrm{Curv}^{\nnabla}\rangle\,.$
\end{enumerate}

Proposition \ref{propsTensCurvTau} guarantees that the graded curvature tensor is graded $\Omega (M)$-multilinear, 
and so, only the first mapping is left $\Omega(M)-$linear. Due to this, we define the graded Ricci tensor as the 
mapping\footnote{Sometimes,
for typographical reasons, we will use the same notation with $\nnabla$ as a subindex.}
$ {\rm Ric^{\nnabla}}:\mathcal X^G(M) \times \mathcal X^G(M)\to \Omega(M)$
given by
 \[
 \langle D_1,D_2;{\rm Ric^{\nnabla}}\rangle:= {\rm STr}\left( D\mapsto \langle D,D_1,D_2;\mathrm{Curv}^{\nnabla}\rangle\right).
 \] 

Explicitly, if $\{E_1,...,E_n;\tilde E_{1},...,\tilde E_{n}\}$ is a homogeneous basis of supervector fields such that
$|E_i|=0$ and $|\tilde{E}_i|=1$ for $i\in\{1,\ldots,n\}$ (we call this a \emph{pure} basis), and
$\{E^{1*},...,E^{n*};\tilde E^{1*},...,\tilde E^{n*}\}$ is the corresponding dual basis as the $\Omega(M)-$linear 
morphism $D\mapsto \langle D,D_1,D_2;\mathrm{Curv}^{\nnabla}\rangle$ is homogeneous with degree $|D_1| + |D_2|$, we get
\begin{equation}\label{Ric_GDual}
 \langle D_1,D_2;{\rm Ric^{\nnabla}}\rangle
           = \sum_{i=1}^p \langle \, \langle E_i, D_1,D_2;\mathrm{Curv}^{\nnabla} \rangle \,; \, E^{i*}\rangle -(-1)^{\left(|D_1|+|D_2|\right)} \sum_{l=p+1}^{p+q}\langle \, \langle \tilde E_l, D_1,D_2;\mathrm{Curv}^{\nnabla} \rangle \,; \, \tilde E^{l*}\rangle\,.        
\end{equation}

If $H:\mathcal X^G(M) \to \mathcal X^G(M)$ is a homogeneous left $\Omega(M))-$linear mapping, then 
\[
{\rm STr} \,H = {\rm STr}\,(H)_{\mathcal B}\,
\] 
where $(H)_{\bf \mathcal B}$ is the supermatrix associated to $H$ in the basis ${\bf \mathcal B}$. 
In particular, if $H$ is the mapping $H:D\mapsto \langle D,D_1,D_2;\mathrm{Curv}^{\nnabla}\rangle$, we can compute
${\rm Ric^{\nnabla}}$ matricially. If, moreover, we have a non-degenerate graded bilinear form $\bm\tau$, we can
give another matrix expression for ${\rm Ric^{\nnabla}}$, this time in terms of the supermatrix associated to
$\bm\tau$ and the graded curvature tensor $R^{\nnabla}_{\bm\tau}$.

\begin{pro}\label{RicMatriz}
Let $\bm\tau$ be a graded non-degenerate covariant $2-$tensor. Fix two supervector fields $D_1,D_2\in\mathcal{X}^G(M)$,
and consider $C(D_1,D_2)$ the left graded $\Omega(M)-$linear mapping given by $D\mapsto \langle D,D_1,D_2;{\rm Curv^{\nnabla}}\rangle$, which is homogeneous with degree $|D_1|+|D_2|$. If ${\bf \mathcal B}=\{E_1,...,E_n;\tilde E_{1},...,\tilde E_{n}\}$ is a pure basis of supervector fields, and ${\bm\tau}$, $(H)_{\mathcal B}$ are the supermatrices
associated to $\bm\tau$ and $H$, respectively, with respect to ${\mathcal B}$, then:

 \begin{equation*}
\langle D_1,D_2;{\rm Ric^{\nnabla}}\rangle  = {\rm STr}_{\bm\tau} \,(C(D_1,D_2))_{\bf \mathcal B} 
                      = {\rm STr}\left\{
  \begin{pmatrix}\,
  \langle E_i,D_1,D_2, E_r  ; R^{\nnabla}_{\bm\tau}\rangle
                    & \langle E_i,D_1,D_2, \tilde E_s ; R^{\nnabla}_{\bm\tau}\rangle \\[5pt]
  \langle \tilde E_j,D_1,D_2, E_r  ; R^{\nnabla}_{\bm\tau}\rangle
                    & \langle \tilde E_j,D_1,D_2, \tilde E_s ; R^{\nnabla}_{\bm\tau}\rangle 
  \,\,\end{pmatrix}
 \left(\bm\tau\right)^{-1}_{\mathcal B}
 \right\}\,.
 \end{equation*}
\end{pro}
\begin{pf}
It is completely analogous to the proof in the non-graded case.
\end{pf}

The advantage of this approach lies in the fact that, when $\bm{\bm\tau}$ has a particularly simple form,
some calculations involving the graded Ricci tensor are much easier when done matricially.

Consider now the supermanifold $(M,\Omega(M))$ and $\bm\tau$ a non-degenerate graded covariant $2-$tensor, along with the induced left graded $\Omega(M)-$linear mappings
\[
\rm Ric_{\nnabla}^\flat :\mathcal X_{\rm Gr}(M) \to \mathcal X_{\rm Gr}^*(M)
\mbox{ and }
 \bm\tau^\flat :\mathcal X_{\rm Gr}(M) \to \mathcal X_{\rm Gr}^*(M) 
\]
which are homogeneous with respective degrees $|\rm Ric_{\nnabla}^\flat|=0$ and $|\bm\tau^\flat|=|\bm\tau|$.
Because of the non-degeneracy of $\bm\tau$, the homogeneous mapping $\bm\tau^\flat$ is invertible, so we have
a well-defined graded-homogeneous, left $\Omega(M)-$linear endomorphism with degree $|\bm\tau|$: 
\[ 
\left(\bm\tau^\flat\right)^{-1}\circ \rm Ric_{\nnabla}^\flat:\mathcal X^G(M) \to \mathcal X^G(M)\,.
\]

Let $\bm\tau$ be a graded metric or symplectic form. The $\bm\tau-$scalar supercurvature of the supermanifold
$(M,\Omega(M))$, endowed with a graded connection $\nnabla$, is then defined as
\[
 {\rm Scal}^{\nnabla}_{\bm\tau} := 
 {\rm STr} \, \left[ \left(\bm\tau^\flat\right)^{-1} \circ \, {\rm Ric_{\nnabla}^\flat}\right].
\]

We have the following practical way for computing ${\rm Scal}^{\nnabla}_{\bm\tau}$.

\begin{pro}\label{scalmat}
 If ${\bf \mathcal B}$ is a pure basis of supervector fields, then
\[
 {\rm Scal}^{\nnabla}_{\bm\tau} = {\rm STr} \left\{ \left({\rm Ric_{\nnabla}^\flat}\right)_{\bf \mathcal B}\, \left(\bm\tau^\flat\right)_{\bf \mathcal B}^{-1} \right\}
\]
where the product in the right-hand side is the usual matrix product.
\end{pro}
\begin{pf}
This is just a consequence of the fact that the composition of left $\Omega(M)-$linear mappings corresponds to the product of the associated supermatrices.
\end{pf}

When $\bm\tau =\bm\omega$ is an odd symplectic form, we call ${\rm Scal}^{\nnabla}_{\bm\omega}$ the \emph{odd symplectic
scalar curvature} of the supermanifold $(M,\Omega (M))$ or, taking into account that this is the only case we will consider
in the sequel, simply \emph{scalar supercurvature}.

\section{Fedosov structures on $(M,\Omega(M))$}\label{sec-fedosov}

Let $(M,\mathcal{A})$ be a supermanifold, endowed with a graded symplectic form $\boldsymbol{\omega}$ and a graded symplectic 
connection $\nnabla$ (i.e., such that $\nnabla \boldsymbol{\omega} =0$). 
The triple $((M,\mathcal{A}),\nnabla,\boldsymbol{\omega} )$ is then called a Fedosov supermanifold, by analogy with the usual 
case. 
We now consider this class of supermanifolds, paying particular attention to the setting of an odd symplectic form 
$\boldsymbol{\omega}$ on the Cartan-Koszul supermanifold $(M,\Omega(M))$.

\begin{pro}\label{sf1}
Let $\nabla$ be a symmetric linear connection on the smooth manifold $M$, and let $\bm\omega_H$ the odd $2-$form determined 
by the isomorfism $H:TM \to T^*M$ as in \eqref{bomegah}. Let $\nnabla$ be a graded connection on
$(M,\Omega(M))$, given as in \eqref{nnabla}. Then, $((M,\Omega(M)),\bm\omega_H,\nnabla)$ is a Fedosov supermanifold 
if and only if the following hold
\begin{equation}\label{sf1eq}
          \begin{array}{lll} 
              H(K_3(X,Y),Z) &=& - H(K_3(X,Z),Y)\\[5pt]
              H(K_2(X,Y),Z) &=& - H(Y,L_3(X,Z)) + (\nabla_Y H)(K_3(X,Z),\cdot) - (\nabla_{K_3(X,Z)} H)(Y,\cdot)\\[5pt]
              H(Y,L_2(X,Z)) &=& H(Z,L_2(X,Y))-(\nabla_Y H)(Z,X)+(\nabla_ZH)(Y,X)+(\nabla_{K_2(X, Y)} H)(Z,\cdot)\\[5pt]
                                   &&  - (\nabla_Z H)(K_2(X,Y),\cdot)
                                    +(\nabla_{Y} H)(K_2(X,Z),\cdot) - (\nabla_{K_2(X,Z)} H)(Y,\cdot)\\[5pt]
              H(K_1(X,Y),Z) &=& H(K_1(X,Z),Y)\\[5pt]
              H(K_0(X,Y),Z) &=& -H(Y,L_1(X,Z)) +(\nabla_XH)(Y,Z) +(\nabla_YH)(K_1(X,Z),\cdot) -(\nabla_{K_1(X,Z)}H)(Y,\cdot)\\[5pt]
              H(Y,L_0(X,Z)) &=& H(Z,L_0(X,Y)) - (\nabla_X\nabla_Y H)(Z,\cdot) + (\nabla_{\nabla_XY}H)(Z,\cdot)\\[5pt]
                              && +(\nabla_X\nabla_Z H)(Y\cdot) - (\nabla_{\nabla_XZ}H)(Y,\cdot) + (\nabla_{K_0(X,Y)}H)(Z,\cdot)\\[5pt]
                              && - (\nabla_ZH)(K_0(X,Y),\cdot) + (\nabla_YH)(K_0(X,Z),\cdot) - (\nabla_{K_0(X,Z)}H)(Y,\cdot)           \,,
             \end{array}
\end{equation}
for any $X,Y,Z\in\mathcal{X}(M)$.
\end{pro}
\begin{pf}
In order to study $\nnabla \bm\omega_H$, it suffices to consider the cases

$$\begin{array}{l}
\langle i_Y,i_Z; \nnabla_{i_X}\bm\omega_H \rangle\\      
\langle \nabla_Y,i_Z; \nnabla_{i_X}\bm\omega_H \rangle\\
\langle \nabla_Y,\nabla_Z ; \nnabla_{i_X}\bm\omega_H \rangle

\end{array}
\hspace{3cm}
\begin{array}{l}
\langle i_Y,i_Z ; \nnabla_{\nabla_X}\bm\omega_H \rangle \\
\langle \nabla_Y,i_Z; \nnabla_{\nabla_X}\bm\omega_H \rangle\\
\langle \nabla_Y,\nabla_Z ; \nnabla_{\nabla_X}\bm\omega_H \rangle
\end{array}
$$

Each one gives an equation involving the tensors $K_i,L_i$. For instance, the
condition $\langle \nabla_Y,i_Z; \nnabla_{i_X}\bm\omega_H \rangle =0$ translates into
\begin{align*}
\langle \nabla_Y,i_Z; \nnabla_{i_X}\bm\omega_H \rangle 
&= i_X(\langle \nabla_Y,i_Z;\bm\omega_H\rangle)- \langle \nnabla_{i_X}\nabla_Y,i_Z;\bm\omega_H\rangle
     -(-1)^{(-1)(0)} \langle \nabla_Y,\nnabla_{i_X}i_Z;\bm\omega_H\rangle \\
&= i_X(\langle \nabla_Y,i_Z;\bm\omega_H\rangle)- \langle \nabla_{K_2(X,Y)} + i_{L_2(X,Y)}, i_Z;\bm\omega_H\rangle -  \langle\nabla_Y, \nabla_{K_3(X,Z)} + i_{L_3(X,Z)};\bm\omega_H\rangle\\  
&=0\,. 
\end{align*}

Taking into account \eqref{bomegah}, we get the equivalent equation
\begin{align*}
&i_X(\langle \nabla_Y,i_Z;\bm\omega_H\rangle)- \langle \nabla_{K_2(X,Y)}, i_Z;\bm\omega_H\rangle -  \langle\nabla_Y, \nabla_{K_3(X,Z)};\bm\omega_H\rangle -\langle\nabla_Y, i_{L_3(X,Z)};\bm\omega_H\rangle \\
=& -i_XH(Y)(Z)+H(K_2(X,Y))(Z)-(\nabla_YH)K_3(X,Z)+(\nabla_{K_3(X,Z)}H)Y+H(Y)(L_3(X,Z)) =0 \,.
\end{align*}

If we use the non-degenerate bilinear form associated to $H$ (which we will denote by the same letter to lighten the
notation), given by $H(X,Y)=H(X)(Y)$ for any $X,Y\in\mathcal{X}(M)$, then we can further simplify the above expression to
\[
H(K_2(X,Y)\cdot,Z) = - H(Y,L_3(X,Z)\cdot) + (\nabla_Y H)(K_3(X,Z),\cdot) - (\nabla_{K_3(X,Z)} H)(Y,\cdot)\,.
\]
The remaining cases are proved in an analogous manner.
\end{pf}

In particular, a graded symplectic connection $\nnabla$ is characterized just by six tensors fields
$K_0,K_3,L_0,L_1,L_2,L_3 :\Gamma TM\otimes\Gamma TM\to \Gamma\Lambda T^*M\otimes\Gamma TM$. 
 
\begin{thm}\label{thm1}
Under the hypothesis of the preceding Proposition, $((M,\Omega (M)),\bm\omega_H,\nnabla)$ is a Fedosov supermanifold, 
with $\nnabla$ symmetric, if and only if $\nnabla$ has the form
\begin{equation}\label{ocho_reducida}
\begin{split}
&\nnabla_{\nabla_X}\nabla_Y=\nabla_{\nabla_XY+K_0(X,Y)} + i_{L_0(X,Y)}\\[3pt]
&\nnabla_{\nabla_X}i_Y = i_{\nabla_X Y+L_1(X,Y)}\\[3pt]
&\nnabla_{i_X}\nabla_Y=i_{L_2(X,Y)}\\[3pt]
&\nnabla_{i_X}i_Y =\nabla_{K_3(X,Y)} + i_{L_3(X,Y)}\,,
\end{split}
\end{equation}
for any $X,Y,Z\in\mathcal{X}(M)$.
\end{thm}
\begin{pf}
The graded connection $\nnabla$ can be characterized as in Proposition \ref{sf1}. In particular, for any vector fields
$X,Y,Z\in\mathcal{X}(M)$ we find from the calculation in the proof above,
\begin{equation*}
i_ZH(K_2(X,Y)) = \nabla_Y(H(K_3(X,Z))) - \nabla_{K_3(X,Z)}(H(Y))-H([Y,K_3(X,Z)])- i_{L_3(X,Z)}H(Y)\,,
\end{equation*}
and, from the fourth equation in \eqref{sf1eq},
$ i_Z H(K_1(X,Y)) = i_Y H(K_1(X,Z))$; so, using \eqref{torsionless-connection} we get
\[
 \begin{array}{rcl}
 \vspace{2mm}
 i_{Y} H(K_1(X,Z))&=&  i_{Z} H(K_1(X,Y))\\
                          \vspace{2mm}
                          &=& i_{Z} H(K_2(Y,X))\\
                         \vspace{2mm}
                          &=& \nabla_X(H(K_3(Y,Z))) - \nabla_{K_3(Y,Z)}(H(X)) -H([X,K_3(Y,Z)]) -i_{L_3(Y,Z)}H(X) \\
                    \vspace{2mm}
                          &=& -\nabla_X(H(K_3(Z,Y))) + \nabla_{K_3(Z,Y)}(H(X)) +H([X,K_3(Z,Y)])+ i_{L_3(Z,Y)}H(X) \\
                     \vspace{2mm}
                     &=&-i_{Y}H(K_2(Z,X))\\
                     \vspace{2mm}
                     &=& -i_{Y}H(K_1(X,Z))\,.
 \end{array}
\]
As $H$ is non degenerate, we conclude that $K_1=0$, which in turn implies
$K_2=0$, and the statement follows.
\end{pf}

We can be more specific when there are additional conditions on $H$ and $\nabla$, as we show next. To this end,
suppose $(M,H,\nabla)$ is either a Riemannian or a symplectic manifold, with $H:TM\to T^*M$ the isomorphism induced by
the corresponding non-degenerate $2-$covariant tensor field. Here, $\nabla$ is a symmetric connection, compatible with $H$ in the sense that $\nabla H=0$.

\begin{cor}\label{VariosCeros_omega_H particular}
Under the preceding conditions, $((M,\Omega (M)),\bm\omega_H,\nnabla)$ is a Fedosov supermanifold with $\nnabla$
graded symmetric, if and only if a $\nnabla$ has the form
\begin{equation}\label{ocho_reducida_rie}
         \begin{array}{lcl}
         \nnabla_{\nabla_X} \nabla_Y   &=& \nabla_{\nabla_X Y + K_0(X,Y)} + i_{L_0(X,Y)} \\         
          \nnabla_{\nabla_X} i_Y &=&  i_{\nabla_X Y+L_1(X,Y)} \\         
          \nnabla_{X} \nabla_Y &=&  i_{L_2(X,Y)} \\         
          \nnabla_{X} i_{Y} &=& \nabla_{K_3(X,Y)}\,,\\         
          \end{array}
\end{equation}
where the tensor fields $K_0,K_3,L_0,L_1,L_2 :\Gamma TM\otimes \Gamma TM\to \Gamma \Lambda T^*M\otimes \Gamma TM$ satisfy:
\begin{equation*}
\begin{array}{ll}
\vspace{2mm}
            K_0\mbox{ is symmetric }, K_3\mbox{ is skew-symmetric, }  &\hspace{1cm}  H(K_3(X,Y),Z)=-H(K_3(X,Z),Y)\,,\\
\vspace{2mm}
            L_0(X,Y)=L_0(Y,X)+\mathrm{Curv}^\nabla(X,Y)\,,  & \hspace{1cm} H(K_0(X,Y),Z) = -H(Y,L_1(X,Z))\,,\\
 \vspace{2mm}
             L_1(X,Y) = L_2(Y, X)\,,  & \hspace{1cm}  H(Y,L_0(X,Z)) = H(Z,L_0(X,Y))\,,
\end{array}
\end{equation*}
for any $X,Y,Z\in\mathcal{X}(M)$. 
\end{cor}
\begin{pf}
The hypothesis in the statement and the second condition in Proposition \ref{sf1} together with the fact that
$K_2=0$ in this case (as a consequence of Theorem \ref{thm1}), imply that
\[
H(K_2(X,Y),Z) = - H(Y,L_3(X,Z))\,,
\]
hence $L_3=0$. On the other hand, the third condition in Proposition \ref{sf1} is now a consequence of the others,
concretely of the fifth, which now reads $H(K_0(X,Y),Z)=-H(Y,L_1(X,Z))$, so (using the properties from Theorem \ref{thm21})
\[
 \begin{array}{rcl}
 \vspace{2mm}
  H(Y,L_2(X,Z)) &=& H(Y,L_1(Z,X))\\
  \vspace{2mm}
                     &=& -H(K_0(Z,Y),X) \\
\vspace{2mm}
                     &=& -H(K_0(Y,Z),X) \\
\vspace{2mm}
                     &=& H(Z,L_1(Y,X))\\
                     &=& H(Z,L_2(X,Y))\,,                
 \end{array}
\]
and the statement follows.
\end{pf}

\section{Vanishing odd symplectic scalar curvature}\label{sec4}

In this section we will use some results from graded linear algebra. References for this topic are \cite{Lei80} and
\cite{Fre86}.

We are interested in a Fedosov supermanifold $((M,\Omega (M)),\bm\omega_H,\nnabla)$, with $\nnabla$ a symmetric graded 
connection, $H:TM\to T^*M$ an isomorphism, and $\bm\omega_H$ the odd symplectic superform defined by $H$ as in \eqref{bomegah}.
Once chosen a symmetric connection $\nabla$, not necessarily compatible with $H$, in a basis of homogeneous supervector fields 
$\{\nabla_{X_i},i_{X_i}\}$ the odd supermatrix associated to $\bm\omega_H$ adopts the form:
\begin{equation}\label{smatrizDeOmegaH}
(\bm\omega_H)
 = \begin{pmatrix}
     P & - H \\
      H^t &0
    \end{pmatrix}
\end{equation}
(where, by an abuse of notation, $H$ also denotes the invertible matrix associated to the isomorphism $H$), so: 
\[
(\omega_H^\flat)
 = \begin{pmatrix}
     P^t & -(-1)^1 (H^t)^t \\
     (-H)^t & (-1)^1 0
   \end{pmatrix}
 = \begin{pmatrix}
    P^t & H \\
    -H^t & 0
   \end{pmatrix}.
\]

On the other hand, a block matrix like
\[
\begin{pmatrix}
     A & B \\
     C & 0
    \end{pmatrix}
\] 
with $B$ and $C$ invertibles, is also invertible with inverse
\[ 
\begin{pmatrix}
     0 & C^{-1} \\
     B^{-1} & -B^{-1}AC^{-1}
    \end{pmatrix}\,;
\] 
we therefore get
\[
(\omega_H^\flat)^{-1}
 = \begin{pmatrix}
     0 & (-H^t)^{-1} \\
     H^{-1} & -(H^{-1})P^t(-H^t)^{-1}
   \end{pmatrix}
 =\begin{pmatrix}
     0 & (-H^t)^{-1} \\
     H^{-1} & H^{-1}P^t(H^t)^{-1}   
  \end{pmatrix}\,.
\]

On the other hand, if the even supermatrix associated to the graded Ricci tensor ${\rm Ric}_{\nnabla}$ is
\begin{equation}\label{smatrizDeRic}
\left({\rm Ric}_{\nnabla}\right)
=\begin{pmatrix}
  R & S\\
  T & U
 \end{pmatrix}\,,
\end{equation}
then
\[
\left({\rm Ric}_{\nnabla}^\flat \right)
=\begin{pmatrix}
  R^t & -(-1)^0T^t\\
  S^t & (-1)^0U^t
 \end{pmatrix}
 = \begin{pmatrix}
    R^t & -T^t \\
    S^t & U^t
   \end{pmatrix}\,,
\]
so we can compute the scalar supercurvature by applying Proposition \ref{scalmat}:
\begin{eqnarray}\label{escalar}
{\rm Scal}^{\nnabla}_{\bm\omega_H}
  &=& {\rm STr}\left[ \left({\rm Ric_{\nnabla}^\flat}\right) \left(\omega_H^\flat\right)^{-1} \right] \nonumber\\
  &=& {\rm STr}\left[ \begin{pmatrix}
    R^t & -T^t \\
    S^t & U^t
    \end{pmatrix}  
    \begin{pmatrix}
     0 & (-H^t)^{-1} \\
     H^{-1} & H^{-1}P^t(H^t)^{-1}   
    \end{pmatrix}
    \right] \nonumber\\
    &=& {\rm STr}\begin{pmatrix}
                  -T^t H^{-1} & *\\
                            * &  -S^t(H^t)^{-1} + U^tH^{-1}P^t(H^t)^{-1}
                  \end{pmatrix} \nonumber\\[5pt]
     &=& {\rm Tr}\,\left[-T^t H^{-1}\right]
         -(-1)^{1}\,{\rm Tr}\, \left[ -S^t(H^t)^{-1} + U^tH^{-1}P^t(H^t)^{-1} \right] \nonumber\\[5pt]
     &=&  {\rm Tr}\,\left[-T^t H^{-1}\right]
         + {\rm Tr}\, \left[ -S^t(H^t)^{-1}\right] + {\rm Tr}\, \left[ U^tH^{-1}P^t(H^t)^{-1} \right].
\end{eqnarray}

Thus, in order to compute the scalar supercurvature we must first know the matrix structure of the graded Ricci
tensor which, in turn, requires an analysis of the graded curvature tensor. If $\nnabla$ is given by the tensor fields
$\{K_i,L_i\}$ as in \eqref{nnabla}, recall that the graded curvature tensor is determined by
\eqref{curvatura_nnabla_ocho}.

\textbf{Important remark}: In the remainder of this section, we will assume that $(M,H,\nabla)$ is either a Riemannian or 
symplectic manifold, with $\nabla$ compatible with $H$, so $\nabla H=0$.

With this assumption the block $P$ in \eqref{smatrizDeOmegaH} vanishes, and then \eqref{escalar} implies that the
scalar supercurvature can be computed as
\begin{equation}\label{scalblocks}
{\rm Scal}^{\nnabla}_{\bm\omega_H} =  {\rm Tr}\,\left[-T^t H^{-1}\right] + {\rm Tr}\, \left[ -S^t(H^t)^{-1}\right].
\end{equation}

If we now take into account that $\nnabla$ is graded symmetric, Corollary \ref{VariosCeros_omega_H particular} 
implies $K_1=K_2=L_3=0$, and the curvature of $\nnabla$ still will be
determined by \eqref{curvatura_nnabla_ocho}, but this time with
\begin{eqnarray}\label{tensores de la curvatura omega_H_gral}
A_1(X,Y,Z)&=& \,\,\,\nabla_X K_0(Y,Z) + K_0(X,\nabla_YZ + K_0(Y,Z))  - K_0([X,Y],Z)+ R^\nabla(X,Y)Z \nonumber\\\vspace{2mm}
           && -\nabla_Y K_0(X,Z)  - K_0(Y,\nabla_XZ + K_0(X,Z)) \nonumber \\            
B_1(X,Y,Z)&=&  \,\,\,\nabla_X L_0(Y,Z) + L_0(X,\nabla_YZ + K_0(Y,Z)) + L_1(X,L_0(Y,Z))  \nonumber\\
           &&  -\nabla_Y L_0(X,Z) - L_0(Y,\nabla_XZ + K_0(X,Z))  -L_1(Y,L_0(X,Z)\nonumber) \\
           &&  - L_0([X,Y],Z)- L_2(R^\nabla(X,Y),Z)\nonumber\\\vspace{2mm}
A_2(X,Y,Z)&=& -K_3(R^\nabla(X,Y),Z)\nonumber\\\vspace{2mm}
B_2(X,Y,Z) &=& \,\,\,\nabla_X L_1(Y,Z) + L_1(X,\nabla_YZ + L_1(Y,Z)) - L_1([X,Y],Z) + R^\nabla(X,Y)Z\nonumber\\
            && -\nabla_Y L_1(X,Z) - L_1(Y,\nabla_XZ + L_1(X,Z)) \nonumber \\ \vspace{2mm}
A_3(X,Y,Z)&=& -K_3(Y,L_0(X,Z))\nonumber\\\vspace{2mm}
B_3(X,Y,Z) &=& L_1(X,L_2(Y,Z)) + \nabla_X L_2(Y,Z) - L_2(Y, \nabla_XZ + K_0(X,Z)) - L_2(\nabla_XY,Z)\nonumber\\\vspace{2mm}
A_4(X,Y,Z)&=& K_0(X,K_3(Y,Z)) + \nabla_X K_3(Y,Z) - K_3(Y,\nabla_XZ+L_1(X,Z))-K_3(\nabla_XY,Z)\nonumber\\\vspace{2mm}
B_4(X,Y,Z) &=& L_0(X,K_3(Y,Z))\nonumber\\\vspace{2mm}
A_5(X,Y,Z) &=& K_3(X,L_2(Y,Z)) + K_3(Y,L_2(X,Z))\nonumber\\\vspace{2mm}
B_5(X,Y,Z) &=& 0 \,=\, A_6(X,Y,Z) \nonumber\\\vspace{2mm}
B_6(X,Y,Z) &=& L_2(X,K_3(Y,Z)) + L_2(Y,K_3(X,Z))\,, \nonumber
    \end{eqnarray}
for any $X,Y,Z\in\mathcal{X}(M)$.
    
In this way, the graded curvature tensor becomes
\begin{equation}\label{tensorDeCurv_omega_H gral}
 \begin{array}{lll}
 \langle \nabla_X,\nabla_Y,\nabla_Z,\nabla_T \,;\, R^{\nnabla}_{\bm\omega_H}\rangle 
      &=& \langle \nabla_{A_1(X,Y,Z)},\nabla_T;\omega_H\rangle +i_{B_1(X,Y,Z)} H(T)  \\[5pt]
 \langle \nabla_X,\nabla_Y,\nabla_Z,i_T \,;\,R^{\nnabla}_{\bm\omega_H}\rangle       
      &=& -i_{T}H(A_1(X,Y,Z)) 
      =  \langle \nabla_X,\nabla_Y,i_T,\nabla_Z \,;\,R^{\nnabla}_{\bm\omega_H}\rangle\\[5pt]
 \langle \nabla_X,i_Y,\nabla_Z,\nabla_T \,;\, R^{\nnabla}_{\bm\omega_H}\rangle      
      &=& \langle \nabla_{A_3(X,Y,Z)},\nabla_T\rangle +i_{B_3(X,Y,Z)}H(T) 
      =  - \langle i_Y,\nabla_X,\nabla_Z,\nabla_T \,;\, R^{\nnabla}_{\bm\omega_H}\rangle \\[5pt]
 \langle \nabla_X,\nabla_Y,i_Z,i_T;R^{\nnabla}_{\bm\omega_H}\rangle 
      &=& -i_T H(A_2(X,Y,Z)) \\
 \langle \nabla_X,i_Y,\nabla_Z,i_T;R^{\nnabla}_{\bm\omega_H}\rangle 
      &=& -i_T H(A_3(X,Y,Z))
      = \langle \nabla_X,i_Y,i_T,\nabla_Z;R^{\nnabla}_{\bm\omega_H}\rangle\\[5pt]
      &=& -\langle i_Y,\nabla_X,\nabla_Z,i_T;R^{\nnabla}_{\bm\omega_H}\rangle
      = - \langle i_Y,\nabla_X,i_T,\nabla_Z;R^{\nnabla}_{\bm\omega_H}\rangle\\[5pt]
\langle i_X,i_Y,\nabla_Z,\nabla_T; R^{\nnabla}_{\bm\omega_H}\rangle 
      &=&  \langle \nabla_{A_5(X,Y,Z)},\nabla_T;\omega_H \rangle \\[5pt]
 \langle \nabla_X,i_Y,i_Z,i_T;R^{\nnabla}_{\bm\omega_H}\rangle 
      &=& -i_T H(A_4(X,Y,Z))
      =   -\langle i_Y,\nabla_X,i_Z,i_T;R^{\nnabla}_{\bm\omega_H}\rangle\\[5pt]
 \langle i_X,i_Y,\nabla_Z,i_T;R^{\nnabla}_{\bm\omega_H}\rangle 
      &=& -i_T H(A_5(X,Y,Z))
      =   \langle i_X,i_Y,i_T,\nabla_Z;R^{\nnabla}_{\bm\omega_H}\rangle\\[5pt]
 \langle i_X,i_Y,i_Z,i_T;R^{\nnabla}_{\bm\omega_H}\rangle &=& 0\,, 
 \end{array}
 \end{equation}
where $X,Y,Z,T\in\mathcal{X}(M)$ are arbitrary.
Using the expression \eqref{bomegah} for $\bm\omega_H$ in this particular case (with $\nabla H=0$), 
these equations reduce to

\begin{equation}\label{tensorDeCurv_omega_H_particular}
 \begin{array}{lll}
 \langle \nabla_X,\nabla_Y,\nabla_Z,\nabla_T \,;\, R^{\nnabla}_{\bm\omega_H}\rangle 
      &=& H(T,B_1(X,Y,Z))  \\
 \langle \nabla_X,\nabla_Y,\nabla_Z,i_T \,;\,R^{\nnabla}_{\bm\omega_H}\rangle       
      &=& -H(A_1(X,Y,Z),T) 
      =  \langle \nabla_X,\nabla_Y,i_T,\nabla_Z \,;\,R^{\nnabla}_{\bm\omega_H}\rangle\\
 \langle \nabla_X,i_Y,\nabla_Z,\nabla_T \,;\, R^{\nnabla}_{\bm\omega_H}\rangle      
      &=& H(T,B_3(X,Y,Z)) 
      =  - \langle i_Y,\nabla_X,\nabla_Z,\nabla_T \,;\, R^{\nnabla}_{\bm\omega_H}\rangle \\
 \langle \nabla_X,\nabla_Y,i_Z,i_T;R^{\nnabla}_{\bm\omega_H}\rangle 
     &=& -H(A_2(X,Y,Z),T) \\
 \langle \nabla_X,i_Y,\nabla_Z,i_T;R^{\nnabla}_{\bm\omega_H}\rangle 
      &=& -H(A_3(X,Y,Z),T)
      = \langle \nabla_X,i_Y,i_T,\nabla_Z;R^{\nnabla}_{\bm\omega_H}\rangle\\
 \langle i_Y,\nabla_X,\nabla_Z,i_T;R^{\nnabla}_{\bm\omega_H}\rangle &=& H(A_3(X,Y,Z),T)
		= \langle i_Y,\nabla_X,i_T,\nabla_Z;R^{\nnabla}_{\bm\omega_H}\rangle\\
\langle i_X,i_Y,\nabla_Z,\nabla_T;R^{\nnabla}_{\bm\omega_H}\rangle 
      &=& 0 \\
 \langle \nabla_X,i_Y,i_Z,i_T;R^{\nnabla}_{\bm\omega_H}\rangle 
      &=& -H(A_4(X,Y,Z),T)
      =   -\langle i_Y,\nabla_X,i_Z,i_T;R^{\nnabla}_{\bm\omega_H}\rangle\\
 \langle i_X,i_Y,\nabla_Z,i_T;R^{\nnabla}_{\bm\omega_H}\rangle 
      &=& -H(A_5(X,Y,Z),T)
      =   \langle i_X,i_Y,i_T,\nabla_Z;R^{\nnabla}_{\bm\omega_H}\rangle\\
 \langle i_X,i_Y,i_Z,i_T;R^{\nnabla}_{\bm\omega_H}\rangle 
	&=& 0 \,.
 \end{array}
 \end{equation}

Now we can study the graded Ricci tensor. We will need the following properties.
\begin{lema}\label{ELlema}
 Under the preceding assumptions, if $\nnabla$ is graded symmetric and symplectic the following hold:
 \begin{enumerate}
  \item $A_3(X,Z,T) = A_3(T,Z,X)-A_2(X,T,Z)\,.$
  \item\label{item2lemma} $H(A_3(X,Z,T),Y) = H(A_3(T,Z,X),Y) -H(A_2(T,X,Y),Z)\,.$
  \item\label{item3lemma} $H(A_3(Z,T,X),Y) = -H(A_3(Z,Y,X),T)\,,$
  \end{enumerate}
for any $X,Y,Z,T\in\mathcal{X}(M)$.
\end{lema}

\begin{pf}\mbox{}
 \begin{enumerate} 
  \item Recall that $L_0(T,X)=L_0(X,T) +R^\nabla(X,T)$ and $K_3$ is skew-symmetric, so the definitions of $A_3$ and $A_2$
  imply:
  \begin{eqnarray*}
         A_3(X,Z,T) &=& -K_3(Z, L_0(X,T)) \\
                    &=& -K_3(Z, L_0(T,X)+ R^\nabla(X,T))\\
                    &=& -K_3(Z, L_0(T,X)) -K_3(Z,R^\nabla(X,T))\\
                    &=& -K_3(Z, L_0(T,X)) +K_3(R^\nabla(X,T),Z)\\
                    &=& A_3(T,Z,X) -A_2(X,T,Z)\,.
        \end{eqnarray*}
   \item On the one hand, applying $H(\cdot,Y)$ to both sides of the equation in the preceding item, we get
   \[
   H(A_3(X,Z,T),Y) = H(A_3(T,Z,X),Y)-H(A_2(X,T,Z),Y).
   \]
   On the other hand, as $\nnabla\omega_H=0$ implies
   $H(K_3(P,Z),Y)=-H(K_3(P,Y),Z)$, and $\mathrm{Curv}^\nabla$ is skew-symmetric,  
   we can compute
   \begin{eqnarray*}
    H(A_2(X,T,Z),Y) &=& H(-K_3(R^\nabla(X,T),Z),Y)\\
                    &=& H(K_3(R^\nabla(T,X),Z),Y)\\
                    &=& -H(K_3(R^\nabla(T,X),Y),Z)\\
                    &=& H(A_2(T,X,Y),Z)\,,
   \end{eqnarray*}
   so
   \begin{eqnarray*}
    H(A_3(X,Z,T),Y) &=& H(A_3(T,Z,X),Y)- H(A_2(X,T,Z),Y)\\
                    &=& H(A_3(T,Z,X),Y) -H(A_2(T,X,Y),Z)\,.\\
   \end{eqnarray*}
   \item It is a direct computation:
   \begin{eqnarray*}
   H(A_3(Z,T,X) ,Y) &=& H(-K_3(T,L_0(Z,X)) , Y) \\
                       &=& H(K_3(L_0(Z,X),T) , Y) \\
                       &=& -H(K_3(L_0(Z,X),Y) , T) \\
                       &=& H( K_3(Y,L_0(Z,X)) , T ) \\
                       &=& - H(A_3 (Z,Y,X),T))\,.
\end{eqnarray*}
 \end{enumerate}
\end{pf}

This Lemma will allow us to prove that whenever $H$ is a Riemannian metric or a symplectic form, we have $T=-S^t$ in \eqref{smatrizDeRic}, directly leading to a vanishing scalar supercurvature.

\begin{pro}\label{Ric_anti}
With the preceding conditions and notations, if $H$ is a Riemannian metric or a symplectic form, then
\[
\langle \nabla_X,i_Y ; {\rm Ric}^{\nnabla} \rangle = - \langle i_Y,\nabla_X; {\rm Ric}^{\nnabla}\rangle\,.
\]
\end{pro}
\begin{dem}
We will detail the case of $H=g^\flat$ a Riemannian (pseudo)metric (so $\nabla$ is the Levi-Civit\`a connection),
the symplectic case is completely analogous.

Take $\{X_i\}$ a local $g-$orthonormal frame, and consider the induced basis of supervector fields 
${\mathcal B}=\{\nabla_{X_i};i_{X_i}\}_{i=1}^n$ in the Fedosov supermanifold $((M,\Omega(M)),\bm\omega_{g},\nnabla)$. 
The dual basis to ${\mathcal B}$ is given by the set of $1-$superforms ${\mathcal B^*}=\{\nabla^*_{X_i};i^*_{X_i}\}$
defined by 
\[
\nabla^*_{X_i}= -i_{i_{X_i}} \bm\omega_{g} 
, \hspace{.5cm}
i^*_{X_i} = i_{\nabla_{X_i}} \bm\omega_{g},
\]
as it is immediate from \eqref{bomegah} for $H=g^\flat$ in this particular case (with $\nabla H=0$), and the fact
that $\{X_i\}$ is $g-$orthonormal.
Thus:
\begin{eqnarray*}
 \langle D_1,D_2;{\rm Ric}^{\nnabla}\rangle 
           &=& {\rm STr}_{\bm\omega_g}\left( D\mapsto \langle D,D_1,D_2;R^{\nnabla}_{\omega_g}\rangle\right)\nonumber\\
           &=& \sum_{i=1}^n \langle \, \langle \nabla_{X_i}, D_1,D_2;R^{\nnabla}_{\omega_g} \rangle \,; \, \nabla^*_{X_i}\rangle  -(-1)^{\left(|D_1|+|D_2|\right)} \sum_{i=1}^{n} \langle \, \langle i_{X_i}, D_1,D_2;R^{\nnabla}_{\omega_g} \rangle \,; \, i^*_{X_i}\rangle \\
           &=&\sum_{i=1}^n -\langle \, \langle \nabla_{X_i}, D_1,D_2;R^{\nnabla}_{\omega_g} \rangle \,, \, i_{X_i} \, ; \, \omega_{g}\rangle -(-1)^{\left(|D_1|+|D_2|\right)} \sum_{i=1}^{n}  \langle \, \langle i_{X_i}, D_1,D_2;R^{\nnabla}_{\omega_g} \rangle \,, \, \nabla_{X_i} \,;\, \omega_{g}\rangle  \nonumber\\
           &=& \sum_{i=1}^n   -\langle \nabla_{X_i}, D_1,D_2, i_{X_i} \, ; \, R^{\nnabla}_{\omega_g} \rangle -(-1)^{\left(|D_1|+|D_2|\right)} \sum_{i=1}^{n} \langle i_{X_i}, D_1,D_2, \nabla_{X_i} \,;\,  R^{\nnabla}_{\omega_g}\rangle.
\end{eqnarray*}

From equations \eqref{tensorDeCurv_omega_H_particular}, determining the graded curvature tensor, we have,
for any $X,Y\in\mathcal{X}(M)$,
 \begin{equation*}
 \langle i_Y,\nabla_X ; {\rm Ric}^{\nnabla} \rangle
     =  \sum_{i=1}^n   g(A_3(X_i,Y,X)\, ,\, X_i)\,,
 \end{equation*}
 and
 \begin{equation*}
 \langle \nabla_X,i_Y ; {\rm Ric}^{\nnabla} \rangle
   =  \sum_{i=1}^n  g(A_2(X_i,X,Y)\, ,\, X_i) 
   +  \sum_{i=1}^{n}   g(  A_3(X,X_i,X_i)\,,\, Y) \,,
 \end{equation*}
 but, from item \ref{item2lemma} in the preceding lemma, 
\[
 g(  A_3(X,X_i,X_i)\,,\, Y) = g(A_3(X_i,X_i,X),Y) - g(A_2(X_i,X,Y),X_i)\,,
\]
hence
 \begin{eqnarray*}
 \langle \nabla_X,i_Y ; {\rm Ric}^{\nnabla}\rangle
   &=&  \sum_{i=1}^n  g(A_2(X_i,X,Y)\, ,\, X_i) \ + \sum_{i=1}^{n} g(A_3(X_i,X_i,X),Y) - g(A_2(X_i,X,Y),X_i) \\
    &=& \sum_{i=1}^{n} g(A_3(X_i,X_i,X),Y) \,.
 \end{eqnarray*}

Finally, an application of item \ref{item3lemma} in the preceding lemma gives:
 \begin{eqnarray*}
 \langle \nabla_X,i_Y ;{\rm Ric}^{\nnabla}\rangle
    &=& \sum_{i=1}^{n}  -g(A_3(X_i,Y,X),X_i) \,,
 \end{eqnarray*}
so $\langle \nabla_X,i_Y;{\rm Ric}^{\nnabla}\rangle = -\langle i_Y,\nabla_X;{\rm Ric}^{\nnabla}\rangle$, which proves the statement.
\end{dem}

This proposition implies that, whenever $H$ represents a (pseudo)Riemannian metric or a symplectic form, we get $T=-S^t$ and
then
\begin{equation}\label{lastscal}
{\rm Scal}^{\nnabla}_{\bm\omega_H}= {\rm Tr}\,\left[-T^t H^{-1}\right]
         + {\rm Tr}\, \left[ T(H^t)^{-1}\right]\,.
\end{equation}

\begin{teo}
If $((M,\Omega(M),\bm\omega_H,\nnabla)$ is a Fedosov supermanifold, with $\nnabla$ graded symmetric,
such that the underlying manifold $(M,H,\nabla)$
is either (pseudo)Riemannian or symplectic with $\nabla H=0$, then
 \[
 {\rm Scal}^{\nnabla}_{\bm\omega_H}= 0.
 \]
\end{teo}
\begin{pf}
Recall from graded linear algebra that any invertible homogeneous block $A$ has the property
\[
(A^t)^{-1}= (-1)^{|A|}(A^{-1})^t\,,
\] 
and, moreover,
\[
{\rm Tr}\,[A^tB] = {\rm Tr}\,[AB^t]\,.
\]
Thus, $H$ being an even homogeneous block, we have $(H^t)^{-1}= (H^{-1})^t$ and substituting in \eqref{lastscal}
we get
\begin{equation*}
{\rm Scal}^{\nnabla}_{\bm\omega_H}  = - {\rm Tr}\,\left[T (H^{-1})^t\right] + {\rm Tr}\, \left[ T(H^{-1})^t\right]=0\,,
\end{equation*}
which proves the statement.
\end{pf}

\section{The case of non-symmetric graded connections}\label{sec5}

We will fix, in the remainder of this section, a symplectic form $w$ on the manifold $M$ (whose dimension will be
$2n$) with its associated skew-symmetric isomorphism $H:T^*M\to TM$, and consider the associated odd
symplectic form $\bomega_H$ on $(M,\Omega(M))$. Also, $\nnabla$ will denote a graded connection such that 
$\nnabla\bm\omega_H=0$.

The calculations in the preceding section show, in retrospect, that the vanishing of the odd symplectic
scalar curvature there was due to the following facts:
\begin{enumerate}
\item The block $P$ in the local expression \eqref{smatrizDeOmegaH} vanishes.
This, in turn, is a consequence of the choice of a parallel isomorphism $\nabla H=0$. 
\item The graded connection $\nnabla$ was taken graded symmetric. In this case, the particular structure of these connections,
described in Corollary \ref{VariosCeros_omega_H particular}, leads to an expression for the graded Ricci tensor which only
depends on $A_3$ (in the notation of Proposition \ref{pro24}), and hence to a vanishing scalar curvature.
\end{enumerate}

We want to keep the first property as it simplifies explicit calculations a lot; thus, assuming $\nabla H=0$,
we need to explore the setting
of non-symmetric graded connections in order to give explicit examples of non-trivial odd symplectic
scalar curvatures. To this end, we will consider a particular family of graded connections parametrized by a single
tensor field $L:\Gamma TM\otimes \Gamma TM\to \Gamma T^*M\otimes \Gamma TM$, whose elements are of the form
\begin{equation}\label{Lconn}
\begin{split}
&\nnabla_{\nabla_X}\nabla_Y=\nabla_{\nabla_XY} + i_{L(X,Y)} \\[3pt]
&\nnabla_{\nabla_X}i_Y = \nabla_{L(X,Y)}+i_{\nabla_X Y} \\[3pt]
&\nnabla_{i_X}\nabla_Y=0 \\[3pt]
&\nnabla_{i_X}i_Y = 0\,.
\end{split}
\end{equation}

Notice that this is a graded connection of even $\mathbb{Z}_2-$degree, but mixed $\mathbb{Z}-$degree ($0$ and $2$).
Also, according to Corollary \ref{VariosCeros_omega_H particular}, it is non-symmetric (due to the presence of the term
$\nabla_{L(X,Y)}$ in $\nnabla_{\nabla_X}i_Y$).

In order for $((M,\Omega(M)),\nnabla,\bm\omega_H)$ to be a Fedosov supermanifold, the conditions in Proposition \ref{sf1} must be satisfied. These reduce themselves (because all the involved tensors are trivial except $K_1=L=L_0$) to the condition on 
$L:\Gamma TM\otimes \Gamma TM\to \Gamma T^*M\otimes \Gamma TM$,
\[
H(L(X,Y),Z)=H(L(X,Z),Y)\,,
\]
which, $H$ being skew-symmetric, can be rewritten as
\begin{equation}\label{Lcond}
H(L(X,Y),Z)+H(Y,L(X,Z))=0\,.
\end{equation}

From the preceding section, we know that the odd symplectic scalar curvature is given by \eqref{scalblocks}, so we
must determine the blocks $S$ and $T$ of the graded Ricci tensor.

As a previous step, we need
the graded curvature $\mathrm{Curv}^{\nnabla}$ of the connection \eqref{Lconn}, which can be computed as in Proposition \ref{pro24}.
The result is
\begin{equation}\label{nonsym1}
 \begin{split}
 \vspace{2mm}
 &\mathrm{Curv}^{\nnabla}(\nabla_X,\nabla_Y)\nabla_Z = \nabla_{A_1(X,Y,Z)} + i_{B_1(X,Y,Z)}  \\
 \vspace{2mm}
 &\mathrm{Curv}^{\nnabla}(\nabla_X,\nabla_Y)i_Z = \nabla_{A_2(X,Y,Z)} + i_{B_2(X,Y,Z)}  \\
 \vspace{2mm}
 &\mathrm{Curv}^{\nnabla}(\nabla_X,i_Y)\nabla_Z  = \nabla_{A_3(X,Y,Z)} + i_{B_3(X,Y,Z)}  
                                             = -\mathrm{Curv}^{\nnabla}(i_Y, \nabla_X)\nabla_Z \\
 \vspace{2mm}
 &\mathrm{Curv}^{\nnabla}(\nabla_X,i_Y)i_Z  = \nabla_{A_4(X,Y,Z)} + i_{B_4(X,Y,Z)}  
                                                  = -\mathrm{Curv}^{\nnabla}( i_Y, \nabla_X)i_Z \\ 
 \vspace{2mm}
 &\mathrm{Curv}^{\nnabla}(i_X,i_Y)\nabla_Z  = \nabla_{A_5(X,Y,Z)} +i_{B_5(X,Y,Z)}\\ 
 &\mathrm{Curv}^{\nnabla}(i_X,i_Y)i_Z = \nabla_{A_6(X,Y,Z)} +i_{B_6(X,Y,Z)}\,,
 \end{split}
 \end{equation}
where, for $X,Y,Z\in\mathcal{X}(M)$ arbitrary,
\begin{align}\label{nonsym2}
&A_1(X,Y,Z)=\mathrm{Curv}^\nabla (X,Y)Z+L(X,L(Y,Z))-L(Y,L(X,Z))= B_2(X,Y,Z) \nonumber\\
&B_1(X,Y,Z)=(\nabla_X L)(Y,Z)-(\nabla_Y L)(X,Z) =A_2(X,Y,Z) 
\end{align}
and the remaining $A_j,B_j=0$ (with $j\in\{3,4,5,6\}$).

Let $\{e_i,f_i\}_{i=1}^n$ be a symplectic frame for the Fedosov (ordinary) manifold $(M,w,\nabla)$, that is, 
$w(e_i,f_j)=\delta_{ij}$, and consider the pure basis of supervector fields
$\mathcal{B}=\{\nabla_{e_i},\nabla_{f_i}; i_{e_i}, i_{f_i}\}_{i=1}^n$ on the supermanifold $(M,\Omega(M))$. 
Its dual basis is the set of graded $1-$forms $\mathcal{B}^*=\{\nabla^*_{e_i},\nabla^*_{f_i}; i^*_{e_i}, i^*_{f_i} \}$ defined
by
\[
\nabla^*_{e_i}=-i_{i_{f_i}} \omega_{H} 
, \hspace{.5cm}
\nabla^*_{f_i} =i_{i_{e_i}} \omega_{H},
\hspace{0.5cm}
i^*_{e_i}= -i_{\nabla_{f_i}} \omega_{H},
\hspace{0.5cm}
i^*_{f_i} =i_{\nabla_{e_i}} \omega_{H}\,.
\]

Therefore,
\begin{eqnarray*}
 \langle D_1,D_2;{\rm Ric^{\bm\omega_H}_{\nnabla}}\rangle 
           &=& {\rm STr}_{\bm\omega_H}\left( D\mapsto \mathrm{Curv}^{\supernnabla}(D,D_1,D_2)\right)\nonumber\\
           &=& \sum_{i=1}^n \left(\langle \,\mathrm{Curv}^{\supernnabla}( \nabla_{e_i}, D_1,D_2)\,; \, \nabla^*_{e_i}\rangle
               + \langle \,   \mathrm{Curv}^{\supernnabla}(\nabla_{f_i}, D_1,D_2) \,; \, \nabla^*_{f_i}\rangle \right)\nonumber\\
           && -(-1)^{\left(|D_1|+|D_2|\right)} \sum_{i=1}^{n} \left( \langle \, \mathrm{Curv}^{\supernnabla}(i_{e_i}, D_1,D_2)  \,; \, i^*_{e_i}\rangle 
               + \langle \, \mathrm{Curv}^{\supernnabla}(i_{f_i}, D_1,D_2)\,; \, i^*_{f_i}\rangle \right) \nonumber\\
           &=& \sum_{i=1}^n \left(   -\langle \, \mathrm{Curv}^{\supernnabla}(\nabla_{e_i}, D_1,D_2) \,, \, i_{f_i} \, ; \, \omega_{H}\rangle
               + \langle \, \mathrm{Curv}^{\supernnabla}(\nabla_{f_i}, D_1,D_2) \,, \, i_{e_i} \, ; \, \omega_{H}\rangle    \right)  \nonumber\\
           && -(-1)^{\left(|D_1|+|D_2|\right)} \sum_{i=1}^{n} \left(    -\langle \, \mathrm{Curv}^{\supernnabla}(i_{e_i}, D_1,D_2) \,, \, \nabla_{f_i} \,;\, \omega_{H}\rangle +\langle \, \mathrm{Curv}^{\supernnabla}(i_{f_i}, D_1,D_2) \,, \, \nabla_{e_i} \,; \, \omega_{H} \rangle \right)\,.
\end{eqnarray*}

Hence, from equations \eqref{nonsym1} describing the curvature $\mathrm{Curv}^{\nnabla}$, 
we get the expression of the $S$ block for the graded Ricci tensor:
 \begin{equation}\label{RicciEjPpal3}
 \langle \nabla_X,i_Y ; {\rm Ric^{\bm\omega_H}_{\nnabla}} \rangle
   =  \sum_{i=1}^n \left( w(A_2(e_i,X,Y)\, ,\, f_i) - w(A_2(f_i,X,Y)\, ,\, e_i) \right) \,,
 \end{equation}
while the $T$ block is always trivial for this particular family of  graded connections, because of the vanishing of $A_3$:
  \begin{equation*}
 \langle i_Y,\nabla_X ; {\rm Ric^{\bm\omega_H}_{\nnabla}} \rangle
     =  \sum_{i=1}^n \left( w(A_3(e_i,Y,X)\, ,\, f_i) - w(A_3(f_i,Y,X)\, ,\, e_i) \right) =0\,.
\end{equation*}

These blocks are to be inserted in \eqref{scalblocks} in order to obtain the odd symplectic scalar curvature of the
Fedosov supermanifold $((M,\Omega(M)),\nnabla,\bm\omega_{w})$.

\section{An example}\label{lastsection}

In this section we present an explicit family of Fedosov structures on the Koszul-Cartan supermanifold over the torus, 
of the form $((T^2,\Omega (T^2)),\nnabla,\bm\omega_{w_0})$ with $\nnabla$ as in \eqref{Lconn} and $w_0$ the canonical 
symplectic structure inherited from $\mathbb{R}^2$ (see Appendix \ref{appendix}), such that their odd symplectic scalar 
curvatures do not vanish. To this end, from these data we must find a tensor field $L$ providing a solution to condition \eqref{Lcond}. As noticed in a previous work \cite{MMV09}, the property that the sympectic curvature on any Fedosov manifold
satisfies:
\[
w(\mathrm{Curv}^\nabla (X,Y)Z,T)=w(\mathrm{Curv}^\nabla (X,Y)T,Z)\,,
\]
means that we can solve \eqref{Lcond} by taking $L$ symmetric (that is $L(X,Y)=L(Y,X)$ for any $X,Y\in\mathcal{X}(M)$ as
vector valued-forms) and defined through
\[
w_0(L(X,Y)Z,T):=w_0(\mathrm{Curv}^\nabla (Z,T)X,Y)\,.
\]

Using \eqref{eqa6} from the Appendix and the notation there \eqref{fgh}, it is immediate to find that
\begin{equation}\label{exprL}
L=\mathrm{Id}\otimes\left(
(4fb+gc)\,\partial^\flat_1 \odot \partial^\flat_1
-4(fa-hd)\,\partial^\flat_1 \odot \partial^\flat_2
-(ga+4hc)\,\partial^\flat_2 \odot \partial^\flat_2
\right)\,,
\end{equation}
where $\mathrm{Id}$ denotes the identity endomorphism on $\mathcal{X}(M)$, and $\odot$ the symmetric tensor product.
This explicit expression allows us to compute
\[
A_2(X,Y,Z)=(\nabla_XL)(Y,Z)-(\nabla_YL)(X,Z)\,,
\]
but some simplifications can be done beforehand. To this end, first notice that for any $X,Y\in\mathcal{X}(M)$,
\begin{equation}\label{exprA}
A_2(X,X,Y)=0\,,
\end{equation}
and then recall from \eqref{scalblocks} that
\[
{\rm Scal}^{\nnabla}_{\bm\omega_H} =  {\rm Tr}\, \left[ -S^t(H^t)^{-1}\right] \,.
\]

As $-(H^t)^{-1}=-H$, if we write
\[
S=\begin{pmatrix}
\langle \nabla_{\partial_1},i_{\partial_1} ; {\rm Ric^{\bm\omega_H}_{\nnabla}} \rangle &
\langle \nabla_{\partial_1},i_{\partial_2} ; {\rm Ric^{\bm\omega_H}_{\nnabla}} \rangle\\
\langle \nabla_{\partial_2},i_{\partial_1} ; {\rm Ric^{\bm\omega_H}_{\nnabla}} \rangle &
\langle \nabla_{\partial_2},i_{\partial_2} ; {\rm Ric^{\bm\omega_H}_{\nnabla}} \rangle
\end{pmatrix}\,,
\]
and make use of \eqref{RicciEjPpal3} and \eqref{exprA}, we see that we only need to compute
\[
{\rm Scal}^{\nnabla}_{\bm\omega_H} =
\langle \nabla_{\partial_2},i_{\partial_1} ; {\rm Ric^{\bm\omega_H}_{\nnabla}} \rangle -
\langle \nabla_{\partial_1},i_{\partial_2} ; {\rm Ric^{\bm\omega_H}_{\nnabla}} \rangle
= w(A_2(\partial_1,\partial_2,\partial_1),\partial_2) + w(A_2(\partial_2,\partial_1,\partial_2),\partial_1)\,,
\]
which is immediate in view of \eqref{exprL}. The result is the non-trivial $1-$form:
\[
{\rm Scal}^{\nnabla}_{\bm\omega_H} =
(4ab^2-4b^3+4(ab-a^2)c+3(c^2d-bd^2))\,\partial^\flat_1
-(4ab^2+4(b^2-a^2)c-4ac^2+(bc-4c^2)d+(4b-a)d^2)\,\partial^\flat_2\,.
\]

\appendix
\section{Fedosov structures on the torus}\label{appendix}

We follow \cite{GRS98} in this appendix. Consider $\mathbb{R}^2$ with global Euclidean coordinates $(x^1,x^2)$, and let $T^2=\mathbb{R}^2/\mathbb{Z}\times\mathbb{Z}$ 
be the $2-$dimensional torus. On $\mathbb{R}^2$ we have the canonical symplectic structure $w_0$ for which the Euclidean
coordinates are also Darboux. Denoting the canonical basis of tangent vector fields\footnote{In the notation of the preceding
section, $n=1$ with $e_1=\partial_1$ and $f_1=\partial_2$.} by $\partial_i$, $i\in\{1,2\}$, and using
Einstein's summation convention over repeated indices, if $X=X^i\partial_i$, $Y=Y^j\partial_j$ are two vector fields, then
\[
w_0(X,Y)=X^1Y^2-X^2Y^1\,,
\]
equivalently,
\[
w_0=\mathrm{d}x^1\wedge\mathrm{d}x^2\,.
\]

Clearly, this form is invariant under translations $(x^1,x^2)\mapsto (x^1+a,x^2+b)$, that is, is left invariant with respect to
the additive Lie group structure of $\mathbb{R}^2$, and thus descends to the quotient $T^2$. The same argument proves that
a Koszul connection $\nabla$ such that its Christoffel symbols $\Gamma^i_{jk}$ are constants, is left invariant and descends
to $T^2$.

If $X\in\mathcal{X}(T^2)$ is a vector field, the dual $1-$form with respect to a non-degenerate bilinear form $B$ will
be denoted $X^\flat_B =i_XB=B(X,\cdot )$, or simply $X^\flat$ if $B$ is understood. Suppose now that $\nabla$ is a symplectic connection with respect to a symplectic form on $T^2$, $w$, so $\mathrm{Tor}^\nabla =0$ and $\nabla w=0$.
From the first condition we get, for every $X,Y\in\mathcal{X}(T^2)$, $\nabla_XY-\nabla_YX=[X,Y]$, which, expressed in a local
basis $X=\partial_i$, $Y=\partial_j$, gives
\[
\Gamma^k_{ij}\partial_k -\Gamma^k_{ji}\partial_k=[\partial_i,\partial_j]^k=0\,,
\]
that is,
\[
\Gamma^k_{ij}=\Gamma^k_{ji}\,.
\]
Let us introduce the Christoffel symbols of the first kind with respect to the bilinear form $w$ (they appear when
computing the local expression of $(\nabla_{\partial_i}\partial_j)^\flat$):
\[
\Gamma_{kij}=w_{kl}\Gamma^l_{ij}\,,
\]
so, in particular,
\begin{equation}\label{eqa1}
\Gamma_{kij}=w_{kl}\Gamma^l_{ij}=w_{kl}\Gamma^l_{ji}=\Gamma_{kji}\,.
\end{equation}

The second condition $\nabla w=0$ is equivalent to the fact that, for every $X,Y,Z\in\mathcal{X}(T^2)$,
\[
X(w(Y,Z))=w(\nabla_XY)+w(Y,\nabla_XZ)=w(Y,\nabla_XZ)-w(Z,\nabla_XY)\,,
\]
so, putting $w=w_{ab}\mathrm{d}x^a\wedge\mathrm{d}x^b$, $X=\partial_i$, $Y=\partial_j$, $Z=\partial_k$, we get
\[
\partial_iw_{jk}=w_{jl}\Gamma^l_{ik}-w_{kl}\Gamma^l_{ij}=\Gamma_{jik}-\Gamma_{kij}\,.
\]
Consequently, if we do have a Darboux coordinate system (in which $w_{ij}$ are constants),
\begin{equation}\label{eqa2}
\Gamma_{jik}=\Gamma_{kij}\,.
\end{equation}

Relations \eqref{eqa1} and \eqref{eqa2} tell us that, for symplectic connections on the torus,
Darboux coordinates correspond to fully symmetric Christoffel symbols
of the first kind, and vice-versa.

With these preliminaries at hand, a connection $\nabla$ and a symplectic form $w$ on the torus $T^2$ 
will determine a Fedosov structure $(T^2,\nabla,w)$ precisely when there exists a coordinate system (which turns out to
be a Darboux system) in which the Christoffel symbols $\Gamma_{ijk}$ are constants and symmetric with respect to any pair 
of indices. In what follows, we will fix the canonical symplectic $2-$form $w=w_0$, so the Fedosov structures we are
going to consider form a $4-$parameter family
\[
\mathcal{F}=\{(T^2,\nabla,w_0):\nabla\equiv (\Gamma_{111},\Gamma_{112},\Gamma_{122},\Gamma_{222})\}\,.
\]

For notational convenience, we will write
\[
a=\Gamma_{111},b=\Gamma_{112},c=\Gamma_{122},d=\Gamma_{222}\,,
\]
and also
\begin{equation}\label{fgh}
f=ac-b^2,g=ad-bc,h=bd-c^2\,.
\end{equation}

Assume that we have one such Fedosov structures on the torus. From the fact that
\[
w^{-1}_0=-w_0=\begin{pmatrix}
0 & -1 \\
1 & 0
\end{pmatrix}\,,
\]
so (using upper indices for the local expression of the inverse $w^{-1}$) $w^{21}_0=1=-w^{12}_0$, together with
\[
w^{rk}_0\Gamma_{kij}=w^{rk}_0(w_0)_{kl}\Gamma^l_{ij}=\delta^r_l\Gamma^l_{ij}=\Gamma^r_{ij}\,,
\]
we get
\[
\Gamma^1_{11}=w^{11}_0\Gamma_{111}+w^{12}_0\Gamma_{211}=-\Gamma_{211}=-b\,,
\]
and similarly
\begin{equation*}
\Gamma^2_{11}=a,\,
\Gamma^1_{12}=\Gamma^1_{21}=-\Gamma^2_{22}=-c ,\,
\Gamma^2_{12}=\Gamma^1_{21}=b ,\,
\Gamma^1_{22}=-d\,.
\end{equation*}

Consequently, the expressions of the covariant derivatives of basis vectors become
\begin{equation}\label{eqa3}
\begin{split}
&\nabla_{\partial_1}\partial_1= \Gamma^i_{11}\partial_i =-b\partial_1 +a\partial_2\\
&\nabla_{\partial_1}\partial_2=  \Gamma^i_{12}\partial_i =-c\partial_1 +b\partial_2\\
&\nabla_{\partial_2}\partial_1=  \Gamma^i_{21}\partial_i =-c\partial_1 +b\partial_2\\
&\nabla_{\partial_2}\partial_2= \Gamma^i_{22}\partial_i =-d\partial_1 +c\partial_2\,.
\end{split}
\end{equation}

The fact that $\nabla w=0$ implies
\[
\nabla_{\partial_i}\partial^\flat_j=\nabla_{\partial_i}(w(\partial_j,\cdot))=\Gamma^k_{ij}w(\partial_k,\cdot)=
\Gamma^k_{ij}\partial^\flat_k\,,
\]
hence the corresponding equations
\begin{equation}\label{eqa4}
\begin{split}
&\nabla_{\partial_1}\partial^\flat_1= \Gamma^i_{11}\partial^\flat_i =-b\partial^\flat_1 +a\partial^\flat_2\\
&\nabla_{\partial_1}\partial^\flat_2=  \Gamma^i_{12}\partial^\flat_i =-c\partial^\flat_1 +b\partial^\flat_2\\
&\nabla_{\partial_2}\partial^\flat_1=  \Gamma^i_{21}\partial^\flat_i =-c\partial^\flat_1 +b\partial^\flat_2\\
&\nabla_{\partial_2}\partial^\flat_2= \Gamma^i_{22}\partial^\flat_i =-d\partial^\flat_1 +c\partial^\flat_2\,.
\end{split}
\end{equation}

Notice that these imply
\begin{equation}\label{eqa5}
\nabla_{\partial_1}(\partial^\flat_1\wedge\partial^\flat_2)=0=\nabla_{\partial_2}(\partial^\flat_1\wedge\partial^\flat_2)\,.
\end{equation}

The curvature can be computed now from its definition \eqref{eq0}. As it is an $\mathrm{End}(TM)-$valued $2-$form, it suffices
to calculate $\mathrm{Curv}^\nabla (\partial_1,\partial_2)$, and it is immediate that
\[
\mathrm{Curv}^\nabla (\partial_1,\partial_2)\partial_1=
\nabla_{\partial_1}\left(\nabla_{\partial_2}\partial_1\right)-\nabla_{\partial_2}\left(\nabla_{\partial_1}\partial_1\right)=
g\partial_1-2f\partial_2\,.
\]

Also,
\[
\mathrm{Curv}^\nabla (\partial_1,\partial_2)\partial_2=
\nabla_{\partial_1}\left(\nabla_{\partial_2}\partial_2\right)-\nabla_{\partial_2}\left(\nabla_{\partial_1}\partial_2\right)=
2h\partial_1-g\partial_2\,,
\]
so we can write
\begin{equation}\label{eqa6}
\mathrm{Curv}^\nabla =\left[
(2h\partial_1-g\partial_2)\otimes\partial^\flat_1
-(g\partial_1-2f\partial_2)\otimes\partial^\flat_2
\right]\otimes
\partial^\flat_1\wedge\partial^\flat_2\,.
\end{equation}

\end{document}